\documentclass[reprint]{JASA}
\UseRawInputEncoding
\begin{document}

\title[JASA/Bubble curtains]{Bubble curtains for noise mitigation: one vs. two}
\author{Simon Beelen}
\email{s.j.beelen@utwente.nl}
\affiliation{Physics of Fluids Group and Max Planck Center for Complex Fluid Dynamics, J. M. Burgers Centre for Fluid Dynamics, University of Twente, P.O. Box 217, 7500AE Enschede, The Netherlands}

\author{Marten Nijhof}
\author{Christ de Jong}
\affiliation{Acoustics and Underwater Warfare Division, TNO, The Hague, 2509JG, Netherlands}

\author{Leen van Wijngaarden}
\affiliation{Physics of Fluids Group and Max Planck Center for Complex Fluid Dynamics, J. M. Burgers Centre for Fluid Dynamics, University of Twente, P.O. Box 217, 7500AE Enschede, The Netherlands}

\author{Dominik Krug}		
\email{d.j.krug@utwente.nl}
\affiliation{Physics of Fluids Group and Max Planck Center for Complex Fluid Dynamics, J. M. Burgers Centre for Fluid Dynamics, University of Twente, P.O. Box 217, 7500AE Enschede, The Netherlands}
\affiliation{Institute of Aerodynamics, RWTH Aachen University, W\"ullnerstraße 5a, 52062 Aachen,
Germany}

\preprint{Beelen et al., JASA}	

\date{\today}

\begin{abstract}
Bubble curtains are widely used to protect marine life from exposure to noise during offshore construction. However, operating a bubble curtain is costly. Therefore optimizing the acoustic effect of the available air is important. An interesting approach is to split the airflow rate into two separate bubble curtains, rather than one single curtain.

This concept is tested experimentally and numerically. The experiments and the model show an increase in performance of the compressed air when it is split between two manifolds. An increased insertion loss of up to 11dB is measured. This increase in performance is possibly due to the fact that the reflective properties of the bubble curtains are maintained when halving the airflow rate. In effect, by splitting the airflow a second acoustic barrier is added. Additionally, the variations in the bubble curtain performance between individual measurements are shown to be largely caused by temporal variations in the air distribution. The applicability of equivalent fluid models for bubble curtains is discussed, and it is shown that accounting for a gap in the bubble curtain, close to the manifold where the bubble curtain is not fully developed, results in better agreement between the modelled and the measured insertion loss.
\end{abstract}

\maketitle

\section{\label{sec:1} Introduction}

The increasing demand for sustainable energy has led to a growing demand for offshore wind farms. Such installations are typically located in relatively shallow waters ($<50\,\mathrm{m}$), such that they can be bottom founded \citep{guo2022review,wang2010research}. For that reason many coastal waters are being considered for the construction of wind farms, which can for example be seen in \citet{EMODNet}. The main method of securing the wind turbines to the bottom is by driving a monopile into the seabed using impact hammers \citep{merchant2019underwater,musial20192018,tsouvalas2020underwater}. The noise generated during this process significantly affects marine life (\citep{dahl2015underwater,popper2022offshore,hastie2019effects}). For this reason other, less noisy, methods of anchoring the wind turbines to the seabed are being developed \citep{igoe2013investigation,spagnoli2013new}, but pile driving remains the dominant method to date.
Bubble curtains have been proven to be effective in mitigating the negative effects of the noise fields generated during pile driving on marine life and are therefore widely used \citep{nehls2016noise,dahne2017bubble}. However, the operation of a bubble curtain adds significantly to the overall construction expenses. Typically, a separate vessel is required to install and operate the bubble curtain and the associated costs can easily exceed 100.000 euro per pile \citep{strietman2018measures}. Efficient use of bubble curtains is therefore important to reduce the associated operating costs. However, reliable compliance with current (\citep{juretzek2021turning}) and future, possibly frequency-dependent \citep{tougaard2017auditory,stober2019effect}, regulations is at least as high a priority and will remain a challenge. \\

There exists a multitude of acoustical models for predicting the sound emission due to pile driving, for which \citet{tsouvalas2020underwater} presents a comprehensive overview. Although there are analytical relations for estimating the unmitigated emitted sound levels due to pile driving \citep{von2022scaling}, most predictions of the  noise levels during pile driving rely on the finite element method (in section \ref{sec:3} these finite element methods will be discussed in more detail). Validation of the applied models for marine pile driving including a bubble curtain, however, is hindered by the lack of measurement data. Acquiring a complete data set involves full scale measurements at sea and is therefore costly. Moreover, if measurements are carried out the resulting data are often not publicly available (\citep{lippert2016compile}). The nature of the problem also makes it difficult to isolate the effect of the bubble curtain, as the results are also affected by sound transmission through the seabed (\citep[e.g.][]{peng2021study}).\\ 
For the cases where experimental data are available, the results of sound mitigation by bubble curtains can vary drastically. \citet{stein2015hydro} measured the sound exposure levels in different directions surrounding a pile driving site with an active bubble curtain. They found the sound exposure level to vary $\sim 10\,\mathrm{dB}$, which they attribute to the shape of the bubble curtain and the irregular air input. This shows the significance of a proper installation of a bubble curtain for reliable compliance. Similarly, a proper implementation of the bubble curtain in acoustic models is important for \emph{a priori} estimates of the sound levels. It is therefore appropriate to focus on the bubble curtain in order to optimise its implementation and thereby maximise its effectiveness in both real-world applications and models. 
\citet{rustemeier2012underwater} tested different air bubble curtain hoses for their sound mitigating properties in a $10\,\mathrm{m}$ deep lake. They found a hose with a porous membrane, generating very small bubbles, to perform significantly better than all hoses with drilled holes of different sizes and spacings. \citet{chmelnizkij2016schlussbericht} discuss the relevant mechanisms for bubble curtain effectiveness to be the reflection, scattering and damping. \citet{rustemeier2012underwater} implemented a model, including the aforementioned effects, and compared it to his findings. To match the observed damping of the bubble curtain generated by the membrane, a bubble size distribution with way larger bubbles than observed had to be assumed. Possibly \citet{rustemeier2012underwater} encountered the limitation of the so called equivalent fluid model they implemented (the equivalent fluid modelling approach is more generally applied to predict the effective speed of sound in bubble curtains \citep{tsouvalas2020underwater}).
\citet{chmelnizkij2016schlussbericht} point out that most equivalent fluid models, including the ones accounting for damping due to bubble oscillations as the one in \citet{rustemeier2012underwater}, assume a large inter-bubble distance, such that bubble-bubble interaction can be neglected, which for typical void fractions observed in bubble curtains is not the case. The inter-bubble distance becomes more relevant close to and at the resonance frequency as the radiation cross section of the bubbles increase here to significantly surpass their physical size. \citet{feuillade1996attenuation} derived a model which does include a term for the bubble-bubble interaction. However, he also indicate that around the resonance frequency non-linearities become prominent, which are not captured by their model.\\

Bubble curtains are effective acoustic barriers at low frequencies, due to the large mismatch in (specific) acoustic impedance between the water and the bubbly layer \citep[e.g.][]{zhu2023modelling}. Particularly at lower frequencies, which are most relevant in the case of pile driving with typical emission peaks between $\approx 100\,\mathrm{Hz}$ and $\approx 500\, \mathrm{Hz}$ \citep{bellmann2014overview,bailey2010assessing}, reflection, resulting from this impedance mismatch, is the dominant factor in sound mitigation \citep{chmelnizkij2016schlussbericht}. A mismatch in acoustic impedance is, in the case of bubble curtains, a consequence of the lower speed of sound in the bubbly mixture as compared to the speed of sound in the water column. The change in density can be neglected since the void fraction inside the bubble curtain is in the order of $\sim 1\,\%$. At the same time it is well known that the dependence of the speed of sound of a bubbly mixture ($c_m$) on the void fraction ($\beta$) is non-linear \citep{wood1956textbook}, which can be simplified to $c_m\approx \sqrt{p_a/\rho_l\beta}$ with $p_a$ the surrounding pressure and $\rho_l$ the liquid density for in the relevant void fraction range for bubble curtains of $0.01\%<\beta<<100\%$ \citep{wijngaarden1972one}. This non-linearity indicates that the mismatch in acoustic impedance is hardly dependent on the amount of air supplied to the bubble curtain, since the speed of sound remains relatively low anyway. A significantly higher speed of sound (closer to that of pure water) requires such a low air supply rate that for practical applications it would not result in a developed bubble curtain useful for practical applications at all. The independence of the acoustic impedance mismatch on the air flow rate also indicates that splitting up the total air flow rate between two manifolds would result in two acoustic barriers with both close to the same impedance as a single bubble curtain. This approach seems from these considerations promising for increasing the effectiveness of the supplied air. However, quantifying the insertion loss of the acoustic barriers is not trivial. \citet{commander1989linear}, for example, quantify the reflection and transmission coefficient of an incoming sound wave based on the void fraction and bubble size distribution inside a homogeneous bubbly mixture. The model by \citet{commander1989linear} is applicable for low void fractions ($\beta<0.1\,\%$) and has not been tested for the application to bubble curtains. Furthermore the effect of the void fraction distribution within the curtain on its sound mitigating properties is unknown. We will therefore investigate the possible effectiveness of generating a second bubble curtain while using the same combined air supply rate. Thereby creating a "free" second bubble curtain, where the bubble curtains with half the total air flow rate have individually close to the same impedance difference as the bubble curtain supplied with the total air flow rate. Using two bubble curtains is already common practice \citep{bellmann2014overview}. However, their use with a relatively short distance between the curtains and without increasing the combined air supply rate has not yet been investigated. \\ 

The aim of this paper is to investigate whether we can increase the performance of a bubble curtain configuration while using the same amount of air. To do so, the quantities used to measure the performance are introduced in \ref{sec:approach}. The experimental setup, used to test the performance of single and double  bubble curtains, is introduced in section \ref{sec:2}. In addition to measurements we have also verified if the intended effect can be captured by modelling. Details of these models are provided in section \ref{sec:3}.  In section \ref{sec:4} the results of the measurements and the simulations are shown, in section \ref{sec:5} the implications of the results are discussed, and finally in section \ref{sec:6} the main conclusions are presented.

\section{Approach}\label{sec:approach}
In this paper we will focus on the performance of configurations with one or two bubble curtains in relatively close proximity. A configuration performs better if it reduces more, or transmits less sound. This can be represented by the frequency dependent Insertion Loss (IL), which we will present in decidecade bands. The IL can be based on either the Sound Pressure Level (SPL), $L_{P,s}$ in the following, or the Sound Exposure Level (SEL), $L_{E,s}$ from now on. Generally the SPL is used for 'continuous' sound and the SEL is used for 'impulse' sound. The IL is defined as 

\begin{equation}
    \text{IL}=L_{P,s,nc}-L_{P,s,c}
    \label{eq:ILdefSPL}
\end{equation}
 or
\begin{equation}
     \text{IL}=L_{E,s,nc}-L_{E,s,c},
    \label{eq:ILdefSEL}
\end{equation}

\noindent
where the subscripts "nc" and "c" stand for no curtain and curtain, respectively. The subscript "s" stands for sound, it is used to indicate the origin of the signal. The IL is thus the difference in SPL or SEL between having no bubble curtain employed and having an active bubble curtain. Generally the SPL is used for continuous sound sources, whereas the SEL is used for impulsive or transient sources. The SPL is defined as

\begin{equation}
    L_{P,s}=10 \log_{10}{\left(\frac{\int_{f_1}^{f_2}P_s^2(f)df}{p_0^2}\right)}\,\mathrm{dB},
\end{equation}

\noindent
with $p_0=1\,\mathrm{\mu Pa}$ the reference sound pressure and $P_s$ the sound pressure spectrum in the frequency domain (Fourier transform of the pressure in the time domain). $f_1$ and $f_2$ are the lower and upper bound of the considered frequency range. Finally the SEL is defined as

\begin{equation}
    L_{E,s}=10\log_{10}\left(\frac{\int_{t_0}^{t_1}{p_s(t)^2}dt}{E_0}\right)\,\mathrm{dB},
\end{equation}

\noindent
with $E_0=1\,\mathrm{\mu Pas}$, $p_s$ the time signal of the pressure. The time span which is used throughout this paper is defined as $\Delta t= t_1-t_0=3\,\mathrm{s}$. \\

\section{\label{sec:2} Experimental setup}

The experiments reported on in this paper were carried out in the so-called Concept Basin (CB) of the Maritime Research Institute Netherlands (MARIN). The CB is a freshwater tank with a length of $220\,\mathrm{m}$, a width of $4\,\mathrm{m}$ and a depth of $3.6\,\mathrm{m}$. Measurements were taken approximately in the middle of the tank, with the bubble curtains emanating from a pipe at the bottom of the tank. 

\begin{figure}[!ht]
\centering
\includegraphics[width=\columnwidth]{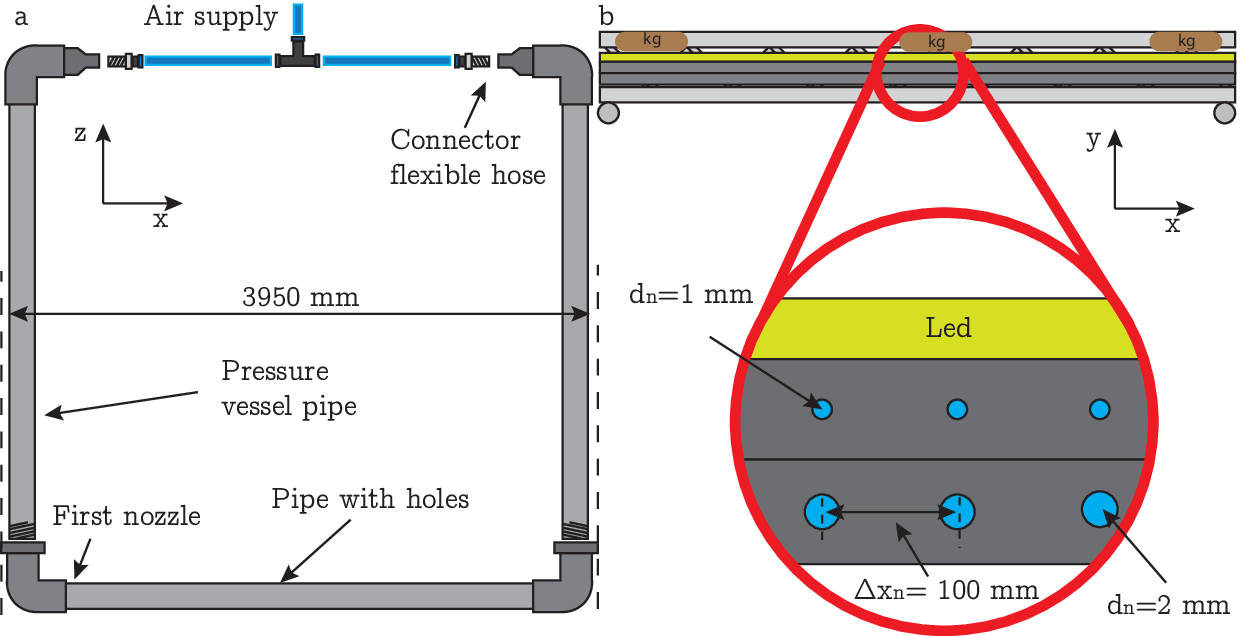}
\caption{(color online) a) Bubble curtain assembly b) Mounting of two bubble curtains on a weighted frame.} \label{fig:Bubble_curtains}
\end{figure}

The bubble curtain(s) were generated from a PVC-pipe with outer diameter $d_{out}=32\,\mathrm{mm}$ and inner diameter $d_{in}=24.8\,\mathrm{mm}$. The large wall thickness of the pipe of $3.6\,\mathrm{mm}$ was one of the measures taken to be able to generate a continuous bubble curtain at low air flow rates since the larger resistance in the nozzle is favourable for an even air distribution along the length of the pipe. In Figure \ref{fig:Bubble_curtains}a a bubble curtain assembly is shown. It consists of the pipe with the holes (the manifold). That pipe is connected via two knee joints to the supply pipes. The first nozzle is drilled close to the knee joint to ensure that the bubble plume covers the entire width up to the basin walls. The distance from the wall to the first nozzle was less then the inter-nozzle spacing ($\Delta x_n=100\,\mathrm{mm}$). The supply pipes are designed to be screwed into the manifold pipe for ease of transport. The total width of the assembly is $3950\,\mathrm{mm}$, to ensure the bubble curtain fits within the basin easily. The air is supplied from two directions into the manifold to reduce the effect of pressure drop over the length of the pipe. The two vertical supply pipes function as a pressure vessel, and reduce the effect of the inflow conditions on the generation of the bubble curtain as compared to directly attaching the air supply to the manifold. Both the pressure vessel effect and the reduced inflow effect contribute to a more continuous bubble curtain particularly at lower air flow rates. The height of the total assembly was approximately $4\,\mathrm{m}$ such that the connections of the supply pipes to the air supply were above the water.

In order to generate bubble curtains with both low and high flow rates we decided to employ two manifolds per bubble curtain location. The first manifold, for the lower airflow rates, has a nozzle diameter of $d_n=1\,\mathrm{mm}$ and the second manifold has a nozzle diameter of $d_n=2\,\mathrm{mm}$. These two were mounted together onto a weighted frame and then lowered to their position on the bottom of the tank. We mounted a LED-strip next to the manifold to visually inspect if the generated bubble curtain was continuous, see Figure \ref{fig:Bubble_curtains}b.\\

\begin{figure}[!ht]
\centering
\includegraphics[width=\columnwidth]{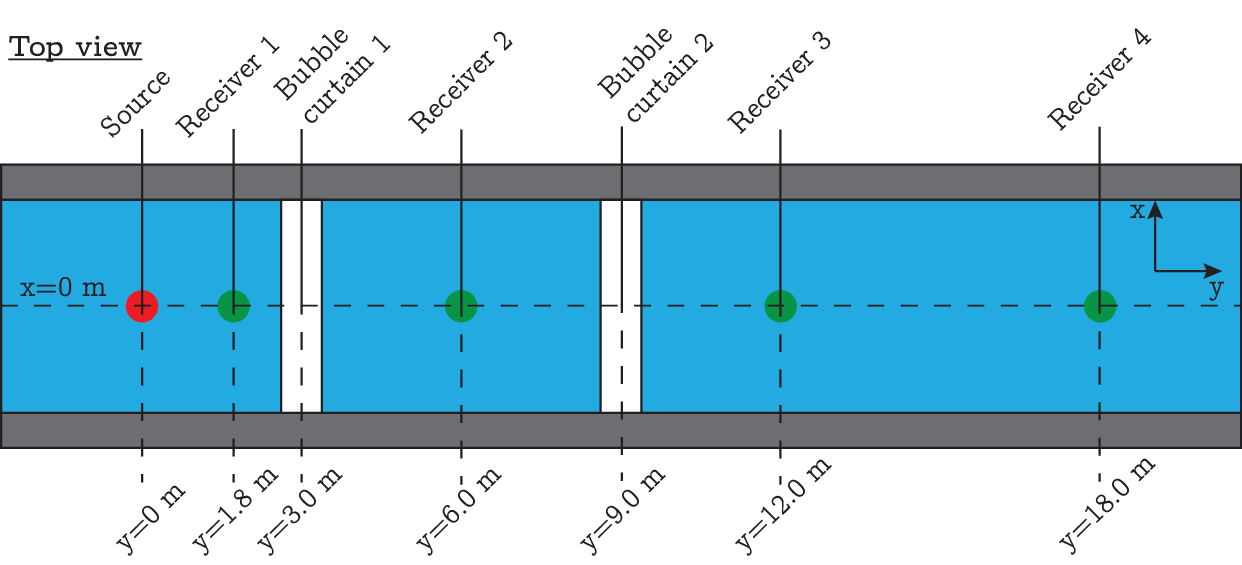}
\caption{(color online) Schematic overview of the setup in the concept basin.} \label{fig:Tankoverview}
\end{figure}

If possible, the air flow rate was controlled by a Bronkhorst F-203AC-FAC-50-V (max flow rate $600\, \mathrm{Lmin^{-1}}$) and otherwise monitored by an Omega FLR-1206 (max flow rate $1400\, \mathrm{Lmin^{-1}}$, max pressure $2.8\,\mathrm{bar}$) and manually controlled using a ball valve. 
The air supply was either through the local pressurized air network at MARIN (up to $600\, \mathrm{Lmin^{-1}}$) or by a diesel air compressor (up to $900\, \mathrm{Lmin^{-1}}$). For flow rates higher than $600\, \mathrm{Lmin^{-1}}$, we generally used the manifolds with the $2\,\mathrm{mm}$ nozzles. To distribute the air to our bubble curtain assemblies we used four ball valves, each of which could manually be opened to supply air to the desired bubble curtain. We only measured the total airflow. This means that if we use two bubble curtains at the same time, we do not know the exact distribution between the two. However, care was taken to ensure that only manifolds with the same nozzle diameters were used, that both ball valves were fully open and that all the supply hoses were of the same length, so that we can assume that the air flow rate is approximately equally distributed to both bubble curtains.\\
\\
In Figure \ref{fig:Tankoverview} the locations of the bubble curtains are shown. Bubble curtain 2 is only employed when we use two bubble curtains. The $x$-coordinate is running parallel to the manifold, the $y$-coordinate is perpendicular to the manifold and the $z$-coordinate points upwards in the vertical direction. The  $x$-coordinate originates in the center of the tank, $y$ at the source location and $z$ at the bottom of the tank. The distance between both bubble curtains is $\Delta y_{bc}=6\,\mathrm{m}$.

Two different sound sources, a J11 hydro sounder/projector produced by USRD (Underwater Sound Reference Division, see \citet{USRD}) and a down-scaled airgun (TNO, The Netherlands) with a volume of $164 \,\mathrm{cm^3}$ and an adjustable pressure ($200$ - $800\,\mathrm{kPa}$), were used. The J11 hydrosounder was driven by a logarithmic sweep between $0.1-10\,\mathrm{kHz}$ in 10 seconds, as shown in appendix \ref{app:appendixAch3}. The J11 also produced some (unintended) higher harmonics. The airgun was primarily used to evaluate the effectiveness of a bubble curtain at one time instance as opposed to the J11 which collects information over longer periods of time. Additionally, it was found to be useful in generating higher sound levels, especially at lower frequencies, in comparison to the J11 hydro sounder. During the measurements, we tuned the pressure of the pulse generated by the airgun such that the first receiver would not clip (overload). Both sound sources were located in the middle of the tank ($x_s=0\,\mathrm{m}$, $y_s=0\,\mathrm{m}$, $z_s=1.8\,\mathrm{m}$) , the sources were never installed simultaneously.\\
The receivers we used were four Brüel \& Kjær 8106 hydrophones, the locations of the receivers are, in accordance with Figure \ref{fig:Tankoverview}: ($x_{r,1}=0\,\mathrm{m}$, $y_{r,1}=1.8\,\mathrm{m}$, $z_{r,1}=1.8\,\mathrm{m}$), ($x_{r,2}=0\,\mathrm{m}$, $y_{r,2}=6.0\,\mathrm{m}$, $z_{r,2}=1.8\,\mathrm{m}$), ($x_{r,3}=0\,\mathrm{m}$, $y_{r,3}=12.0\,\mathrm{m}$, $z_{r,3}=1.8\,\mathrm{m}$) and ($x_{r,4}=0\,\mathrm{m}$, $y_{r,4}=18.0\,\mathrm{m}$, $z_{r,4}=1.8\,\mathrm{m}$).\\
The measurements with the J11 hydro sounder were recorded for 60 seconds, due to a minor delay in the signal this resulted in five fully recorded sweeps. The mini airgun recordings took five seconds, however, only the first three seconds after the trigger were used in our analysis. We took recordings of the background noise levels for 30 seconds at three instances: (1) before the measurements, (2) between the measurements with the J11 hydro sounder and the mini airgun (3) after the measurements with the airgun. Finally we also recorded the noise level of the bubble curtain(s) for 30 seconds directly before or after we carried out the measurements with the source, as to make sure we measured at exactly the same air flow rate. The hydrophones were calibrated using a pistonphone before and after the experiments.

\section{Model\label{sec:3}}
To better understand the measurement results, we employ a model and compare the results with the measurement data. The model we employ is similar to most models used for sound emission predictions including bubble curtains found in literature. Specifically, the use of the equivalent fluid model using e.g. \citet{commander1989linear} (or related equivalent fluid models) is widely adopted in literature.
This approach involves characterizing the bubble curtain by its local speed of sound \citep[e.g.][]{domenico1982acoustic}. This approximation is valid in cases where the acoustic wavelengths are much larger than the typical bubble sizes \citep{wood1956textbook}. We use the same modelling approach as used for marine pile driving, albeit, without modelling the source excitation and without the necessity for a seabed model. Only the insertion loss is considered (forgoing the need to know the source characteristics) and our experiments take place in a lab environment.

The modelled domain throughout this paper corresponds with Figure \ref{fig:Tankoverview} (top view) and is also shown in Figure \ref{fig:Model} (side view). The computational cost of the model is limited by using a so-called 2.5D approach. In this approach the full 3D solution is approximated by summing over so-called `propagating modes'. In this context a propagating mode is defined as the combination of an assumed solution in $x$-direction that fits the boundary conditions on the side of the tank (acoustically hard) and the solution of an associated Finite Element (FE) calculation for the $y$ and $z$-direction. In the FE calculation the wave number, for the combined $y$ and $z$-direction, is the wave number in the medium reduced by the wave number in $x$-direction. Only modes with a real (reduced) wave number for the FE calculation are considered  propagating modes. Modes for which the (reduced) wave number are complex are considered evanescent and decay exponentially in $y$/$z$-direction thus leading to minor contributions to the total solution at larger ranges. 
The solutions of these propagation modes are cosines in the $x$-direction since the source and receiver are located in the center of the basin thus requiring only symmetric modes to be included. Neglecting the evanescent modes reduces the computational costs significantly. The number of included propagation modes in practice depends on the frequency of the sound. The Finite Element method is known to converge to the true solution of the mathematical problem (i.e. the Helmholtz equation) and provides a good approximation for acoustic problems, as long as the mesh is generated carefully (see appendix \ref{app:appendixB}). One of the benefits of the FE method for acoustic applications is that it is able to represent infinitely long wave guides efficiently using Perfectly Matched Layers (PMLs), reducing the computational costs significantly. Moreover implementation of the bubble curtain in the domain is relatively straight forward. The relevant details of the modelling approach will be provided in the following.

\subsection{Frequency content}
The considered frequency range, in terms of the centers of the decidecade bands, is $31.5\,\mathrm{Hz}<f<5000\,\mathrm{Hz}$. Within these decidecade bands we chose to resolve five logarithmically distributed subfrequencies, and the energy/power spectral density in the associated sub-bands is assumed to be constant within these sub-bands. The higher frequencies are more computationally intensive since they require a relatively small mesh size due to the smaller wave lengths. It was therefore chosen to limit our range to $5000\,\mathrm{Hz}$. This is acceptable since typical pile driving sound mainly contains low frequency components ($<2000\,\mathrm{Hz}$ \citep{bellmann2014overview}).  

\subsection{Acoustic representation of a bubble curtain} \label{subsec:Modelcequi}
The bubble curtain is represented as an equivalent fluid by using an equivalent speed of sound $c_{m}$ in the bubbly mixture. The speed of sound in the bubble mixture is a complex number where the imaginary part represents the damping component. To determine $c_{m}$, we employ the well known model of \citet{commander1989linear}, in which the bubble response is linearized. According to this model, the sound speed depends on the frequency $f$, the gas fraction $\epsilon_{g}$ and the bubble size distribution $f_{\mathrm{num}}$. The latter two can be combined into the bubble concentration (number of bubbles of a certain size per unit volume)
\begin{equation}
C(a,x,y,z)=\frac{\epsilon_{g}(x,y,z)}{4/3\pi\int_0^{\infty} a^3f_{\mathrm{num}}(a,x,y,z)da} f_{\mathrm{num}}(a,x,y,z),
\end{equation}

\noindent
where $a$ is the bubble radius undisturbed by the sound pressure. The local void fraction in the bubble plume can be modelled using a planar plume integral model. To that end we use the model derived and validated in our previous work \citep{beelen2024planar}, which is based on the integral equations starting from individual round plumes giving us the local void fraction $\epsilon_{\mathrm{g}}(x,y,z)$. The model also predicts the local bubble size distribution, albeit only as a function of the vertical position above the nozzle ($f_{\mathrm{num}}(a,z)$). This is consistent with the observed limited variation in the bubble size distribution in the spanwise ($y$) direction \citep[e.g.][]{beelen2023situ}. The model thus provides a full bubble concentration profile throughout the bubble curtain. The local speed of sound of the mixture $c_m(f,C)$ is given by

\begin{equation}
    \frac{c^2}{c_m^2}=1+4\pi c^2 \int_0^{\infty}{\frac{aC(a)}{\omega_0^2-\omega^2+2ib\omega}}da,
    \label{eq:localSoS}
\end{equation}

\noindent
where $\omega=2\pi f$ is the radial frequency, $c=1481\,\mathrm{m/s}$ is the speed of sound in water and $i$ is the imaginary unit. The eigenfrequency $\omega_0$ and damping coefficient $b$ of the bubbles in the mixture are defined as

\begin{equation}
    \omega_0^2=\frac{p_0}{\rho a^2}\left(\text{Re}\Phi-\frac{2\sigma}{ap_0}\right)
\end{equation}
and 
\begin{equation}
    b=\frac{2\mu}{\rho a^2}+\frac{p_0}{2\rho a^2 \omega}\text{Im}\Phi+\frac{\omega^2a}{2c},
    \label{eq:dampingconstant}
\end{equation}

\noindent
with the undisturbed pressure in the bubble $p_0=p_{\infty}+2\sigma/a$, $p_{\infty}$ the equilibrium pressure in the surrounding liquid and $\sigma=72.8\,\mathrm{mN/m}$ the surface tension of the bubble. $\rho=1000\,\mathrm{kg/m^3}$ and $\mu=1\,\mathrm{mPas}$ are the density and viscosity of the surrounding water. The complex valued function $\Phi$ describes the heat transport inside the bubble,

\begin{equation}
    \Phi=\frac{3\gamma}{1-3i(\gamma-1)\xi\left(\sqrt{i/\xi}\coth{\sqrt{i/\xi}}-1\right)},
\end{equation}

\noindent
and depends on the specific heat ratio of air $\gamma=1.4$ and $\xi=D/\omega a$ with $D=18.46\cdot 10^{-6} \,\mathrm{m^2/s}$ the thermal diffusivity of air. The damping constant (Eq. \ref{eq:dampingconstant}) comprises three terms representing, from left to right, the viscous damping, damping due to heat dissipation, and damping due to acoustic radiation. \\
For effective implementation it is important to note that the integral in Eq. \ref{eq:localSoS} does not need to be evaluated at every horizontal location. This is because the bubble size distribution is assumed to only vary with the vertical distance from the nozzle in the model. Therefor, the following representation, where the void fraction is outside of the integral, significantly reduces the computational cost of calculating the equivalent speed of sound at many different locations throughout the curtain, which is the required input to the FE model:

\begin{equation}
    \frac{c^2}{c_m^2}=1+\epsilon_{g}4\pi c^2 \int_0^{\infty}{\frac{a f_{\mathrm{num}}/\left(4/3\pi\int_0^{\infty} a^3f_{\mathrm{num}}da\right) }{\omega_0^2-\omega^2+2ib\omega}}da.
    \label{eq:SpeedofSound}
\end{equation}
\noindent
For a given bubble size distribution and void fraction distribution, we retrieve a distribution of the speed of sound in the mixture ($c_{m}(x,y,z)$). As previously discussed, we utilise a 2D FE model to compute the sound propagation. To reduce the 3D situation to a 2D representation we first calculate the average void fraction in the $x$-direction and then compute the corresponding speed of sound. This choice is substantiated in appendix \ref{app:SpeedofsoundAVG}, where the effect of structures (or 'holes') in the void fraction distribution of real bubble curtains is discussed in relation to this choice. We further focus in on the difficulties associated with reducing a 3D situation to the 2D FE domain.

\subsection{Implementation and results}
The FE model is implemented in Comsol Multiphysics 6.0. This commercially available software allows for easy FE implementation of acoustic problems. For the sake of brevity the detailed implementation is discussed in appendix \ref{app:appendixB}. The numerical domain (see Figure \ref{fig:Model} in appendix \ref{app:appendixB}) covers the full experimental configuration in the $yz$-plane.
The speed of sound within the bubble curtain is calculated according to Eq. \ref{eq:SpeedofSound}. The performance of the bubble curtain will be presented in terms of the IL as discussed in section \ref{sec:approach}. The IL is calculated based on the SPL (Eq. \ref{eq:ILdefSPL}) or the SEL (Eq. \ref{eq:ILdefSEL}).

\subsection{The effect of `holes' in the bubble screen} \label{subsec:Transparency}

Close to the manifold the bubble plume is not yet continuous leaving 'holes' in the $x,z$-plane of the bubble curtain. These holes in between the individual nozzles do not contain bubbles and possibly reduce the effectiveness of the bubble curtain. We capture the fraction of `open', i.e. largely void of bubbles, surface area in the bubble curtain in terms of a transparency factor. As shown in appendix \ref{app:SpeedofsoundAVG} there is no direct way of incorporating transparency directly through the effective speed of sound in a 2D FE simulation. To assess the impact of openings in the curtain, at least approximately, we introduce a gap filled with water containing no bubbles at the bottom of the modelled bubble curtain. Based on the approximate distance at which individual plumes merge \citep{beelen2024planar} this gap extends $2\Delta x_n=200\,\mathrm{mm}$ upwards from the manifold and we set $c_m=1481\,\mathrm{ms^{-1}}$ in this region. This approach is in line with scenario 3 of \citet{peng2021study}, who also represent the zone of individual plumes as a leakage region which is characterized by the acoustic properties of water.

\section{Results\label{sec:4}}

In this section we will present the results of the measurements focusing on the IL and compare to model predictions where relevant. As given by Eqs. \ref{eq:ILdefSPL} and \ref{eq:ILdefSEL}, the insertion loss is the difference between the sound levels with and without a bubble curtain. In practice, however, the measured levels may be significantly influenced by the background noise levels. In order to suppress the effect of the background noise on the measured spectrum, we subtract it from our measured signal $P_s^2=P_{s+n}^2-P_n^2$, where $P_{s+n}$ is the measured pressure spectrum resulting from the sound of the source and the background noise and $P_n$ is the background noise pressure spectrum. $P_s$ is the Fourier transformed sound pressure ($\mathrm{Pas}$) due to the source and can be used to determine the IL (Eq. \ref{eq:ILdefSPL}). The IL based on the SEL (Eq. \ref{eq:ILdefSEL}) is corrected for the background noise levels similarly to the SPL and the term $p_s^2$ in the integral is replaced by $p_s^2=p_{s+n}^2-p_n^2$ to correct the SEL for noise. These approximation are accurate when the sound level and the background noise level are uncorrelated. Furthermore, the correction is only meaningful if the sound level of the signal of interest is higher than the sound level of the noise. \\
The background noise levels (shown in appendix \ref{app:backgroundnoise}) are measured when the sound source and bubble curtains are not active. The  maximum recorded background noise (Bg, max in the following figures) across measurements at different times is used as the depicted background noise level in the following. The noise generated by the bubble curtains is also considered background noise. As shown in appendix \ref{app:BCnoise}, the noise level is flow rate dependent. In particular for the $1\,\mathrm{mm}$ nozzle manifold we observe a strong dependence on the airflow rate, whereas the noise is only weakly dependent on the airflow rate when using the $2\,\mathrm{mm}$ nozzles. If the transmitted sound level from the source at a receiver approaches the noise level, the contribution of the source becomes indistinguishable, and the correction for noise proposed above becomes invalid. The source is considered indistinguishable if the total received sound level is less than the $3\,\mathrm{dB}$ higher than the background noise level ($L_{P,s+n}<L_{P,n}+3\,\mathrm{dB}$). When this condition is met we set $L_{P,s}=L_{P,n}$ (similar for SEL) to clearly indicate the signal-to-noise limitation.
Regions where the sound of the source is no longer distinguishable from the noise will also be indicated by a grey area in the following figures.\\
The sound generated by the sources without the bubble curtain active is explicitly used in the definition of the IL ($L_{P,s,nc}$ and $L_{E,s,nc}$ in Eqs. \ref{eq:ILdefSPL} and \ref{eq:ILdefSEL} respectively) and is shown in appendix \ref{app:soundsource} for the fourth receiver. Between different sweeps or different shots of the airgun the produced signal changes slightly. Therefore, we will use the average sound levels of 5 sweeps or 6 shots of the airgun in (most of) the following analysis to determine the IL. Variations in the generated signal are particularly relevant for the airgun as we will present in section \ref{sec:ILchangesch3}. The $3\,\mathrm{s}$ time interval for the SEL is chosen such that it captures the full event of approximately $1\,\mathrm{s}$ with sufficient resolution in the frequency domain, while maintaining a limited influence of noise. At lower frequencies ($<250\,\mathrm{Hz}$) the airgun generates significantly more sound than the hydrosounder and also at higher frequencies the level remains somewhat higher (see appendix \ref{app:soundsource}).\\
In the following we will first show an example of the sound pressure levels used to determine the IL so that we can show the general trend for all measurements and then move to the measured and modelled insertion loss. 

\subsection{Sound pressure levels throughout the measurement domain}

\begin{figure}[!ht]
\centering
\includegraphics[width=\columnwidth]{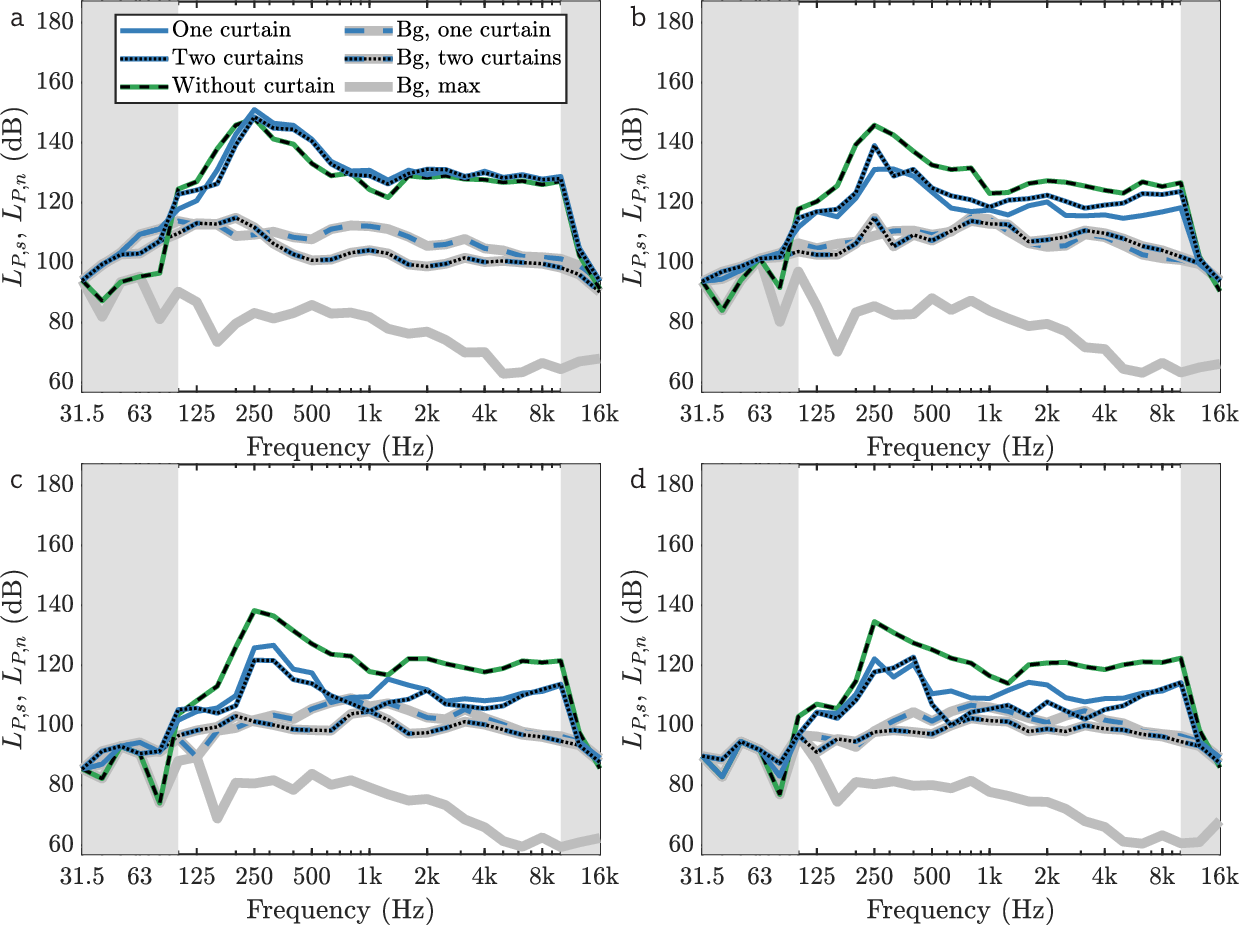}
\caption{(color online) Sound pressure levels for one or two bubble curtains with a total airflow rate of $96\,\mathrm{Lmin^{-1}}$, without a bubble curtain and noise levels at a) receiver 1 b) receiver 2 c) receiver 3 d) receiver 4.} \label{fig:Alllines}
\end{figure}

In Figure \ref{fig:Alllines} we show the sound pressure levels at all four receivers without, with one, and with two bubble curtains for an airflow rate of $96\,\mathrm{Lmin^{-1}}$, measured with and without the hydrosounder. 

The IL at every receiver is given by subtracting the sound level without a bubble curtain (green black dashed lines) from the sound level with a bubble curtain (solid blue line or solid blue line with black dots for one and two curtains respectively). 
When operating a bubble curtain, the overall sound pressure level at the first receiver (Figure \ref{fig:Alllines}a) increases for frequencies exceeding $250\,\mathrm{Hz}$ due to reflection of the sound at the bubble curtain. The second receiver (Figure \ref{fig:Alllines}b) is placed behind the first bubble curtain, which reduces the sound pressure level when it is active. When the second bubble curtain is operated additionally, the sound pressure level between both curtains is seen to increase. The increase is likely due to the reflection of sound at the second curtain and marginally due to reduced effectiveness of the first curtain resulting from the reduced airflow rate. The third receiver (Figure \ref{fig:Alllines}c) is positioned behind both curtains and records lower sound levels overall, as expected, when using two bubble curtains. At the fourth receiver (Figure \ref{fig:Alllines}d) we mostly observe a lower sound pressure level when operating two bubble curtains, similar to the third receiver. The trends discussed here for an airflow rate of $96\,\mathrm{Lmin^{-1}}$ are similarly observed for all airflow rates. However, the bubble curtain noise becomes more problematic for our measurements when the airflow rate is increased and a louder source would be required to ensure a good signal to noise ratio for these cases. This can already be seen in Figure \ref{fig:Alllines}d, where the sound pressure level gets very close to the bubble curtain noise level around $630\,\mathrm{Hz}$. For higher airflow rates the region of the plot where the signal of interest is not above the noise floor extends to more frequencies. So far in this section we have shown a few examples of the sound pressure levels which can be used to calculate the IL. In the next part of this section we will present the measured and modelled IL for different airflow rates and source types.

\subsection{Insertion loss based on hydrosounder measurements} \label{sec:ILchangesch3}

\begin{figure}[!ht]
\centering
\includegraphics[width=\columnwidth]{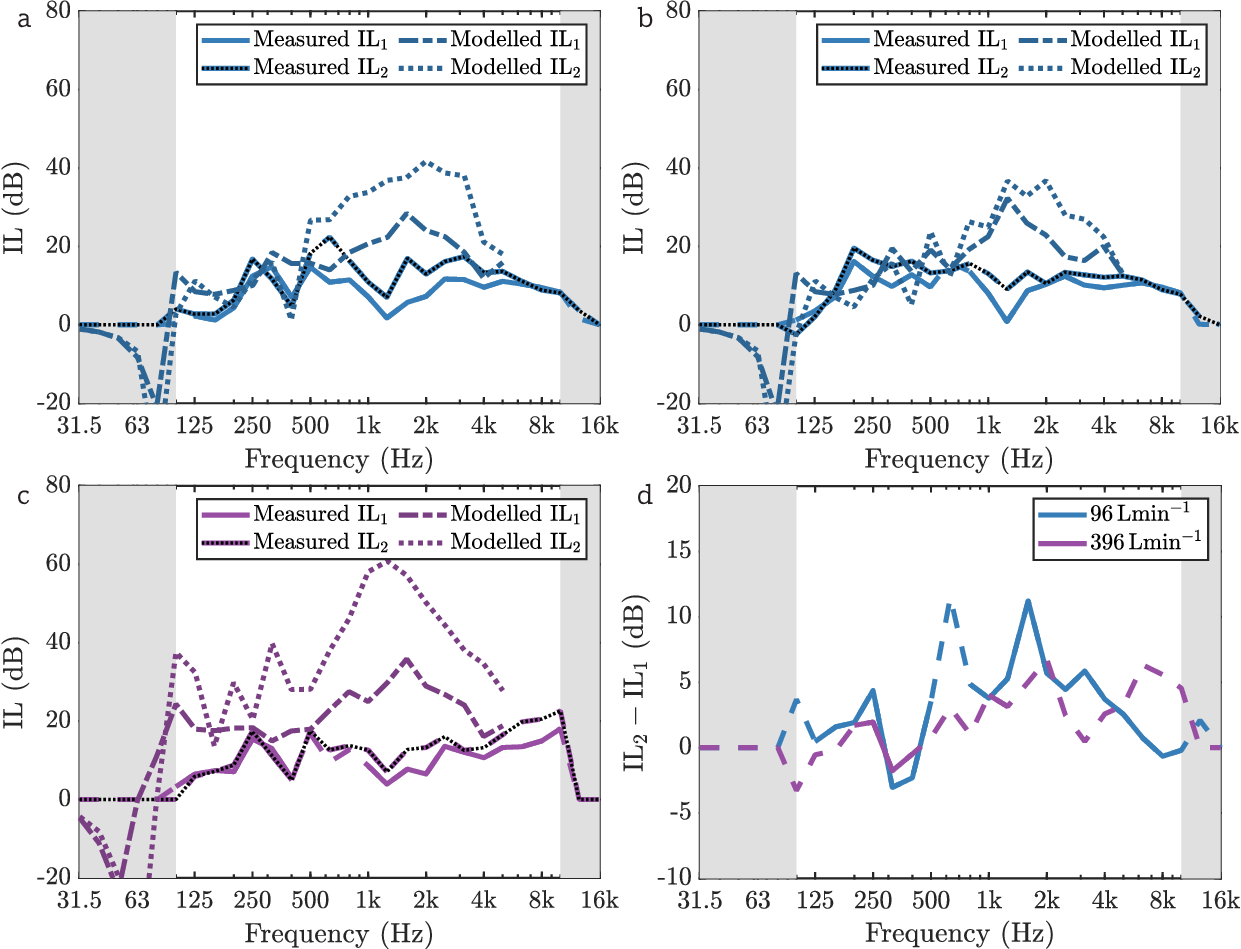}
\caption{(color online) The measured and modelled IL for different configurations with an air flow rate of $96\,\mathrm{Lmin^{-1}}$ at a) receiver 4 and b) receiver 3. c) for an air flow rate of $396\,\mathrm{Lmin^{-1}}$ at receiver 4. d) the measured difference in IL at receiver 4 for one and two bubble curtains.} \label{fig:ILSWEEP}
\end{figure}

We can calculate the IL from the sound levels plotted in Figure \ref{fig:Alllines}. The general trend in Figure \ref{fig:Alllines} is seen for all airflow rates: The IL at the first receiver is negative due to reflection of the sound at the bubble curtain. The IL at the second receiver is positive, i.e. the sound level is reduced, and the value is higher for one bubble curtain than for two. At the third and fourth receivers, the IL is also positive and the two bubble curtain setup outperforms the one bubble curtain setup.\\ 
In Figures \ref{fig:ILSWEEP}a and b we show the IL for receiver 4 and 3 respectively. We will focus mainly on the IL measured at the fourth receiver since it is farthest away from the bubble curtain and we assume the performance of a bubble curtain further away from the source to be more relevant to the real-life application of bubble curtains. However since the IL depends on the receiver location, the IL at both receiver locations 3 and 4 are shown once. \\
Particularly in the range of $0.5-4\,\mathrm{kHz}$ the IL increases significantly. In the relevant frequency range of $0.1-3\,\mathrm{kHz}$ the IL of both the one and two bubble curtain setup varies between 3 and 15 dB and 3 and 20 dB respectively. Further we see a dip in the IL around $1250\,\mathrm{Hz}$ the precise origin of which is unknown. The source level around $1250\,\mathrm{Hz}$ also contains a dip (see Figure \ref{fig:Alllines}) possibly indicating that the source is interacting with a resonance of the basin geometry.\\
For the low airflow ($96\,\mathrm{L/min}$) rate the IL derived from the measurements is hardly influenced by the noise generated by the bubble curtains. For frequencies where the noise influences the IL derived from the measurements we display the IL as a dashed line. In Figure \ref{fig:ILSWEEP}c the IL for an airflow rate of $396\,\mathrm{Lmin^{-1}}$ is shown and, as indicated by the dashed lines, it is for most frequencies limited by the noise generated by the bubble curtain(s). Once the IL is limited by the bubble curtain noise it should be interpreted as a lower bound for the actual IL. A louder source would be required to accurately determine the true IL in these cases. Still the measured values are meaningful in the sense of a minimum IL of the bubble curtain and is therefore shown.

To illustrate the effectiveness of splitting up the airflow rate, the IL of a single bubble curtain configuration is subtracted from the IL of the two bubble curtain configuration, $\mathrm{IL}_2-\mathrm{IL}_1$. This is shown for both $96\,\mathrm{Lmin^{-1}}$ and $396\,\mathrm{Lmin^{-1}}$ in Figure \ref{fig:ILSWEEP}d. For airflow rates higher than $396\,\mathrm{Lmin^{-1}}$ (not shown) both the one and two curtain IL derived from the measurements are limited by the noise floor such that the difference between the two is no longer meaningful and therefore not included. In Figure \ref{fig:ILSWEEP}d the dashed line indicates that the noise floor influenced the results. The difference in IL for the two shown airflow rates is mostly positive, meaning splitting up the air into two distinct bubble curtains increases the effectiveness of the used compressed air. For the frequencies $630\,\mathrm{Hz}$ and $1600\,\mathrm{Hz}$  the increase in IL even exceeds $11\,\mathrm{dB}$ at an airflow rate of $96\,\mathrm{Lmin^{-1}}$. For $396\,\mathrm{Lmin^{-1}}$ the difference is less distinct since the IL of the two bubble curtain setup was limited by the noise floor, nevertheless significant IL increases exceeding 6 dB are observed. For both flow rates the two bubble curtain setup performs worse than the one bubble curtain setup between $250-400\,\mathrm{Hz}$. We will comment on these observations in section \ref{sec:5}.

\subsection{Results for the insertion loss based on the model}
\label{sec:res_model}
Also included in Figures \ref{fig:ILSWEEP}a-c is a comparison between the measured and the modelled IL. This is most informative for the lowest airflow rate (Figures \ref{fig:ILSWEEP}a and b), since the noise generated by the bubble curtain does not affect the measurement of the IL in this case. For frequencies between $500\,\mathrm{Hz}$ and $3150\,\mathrm{Hz}$ the IL of a single bubble curtain is overpredicted significantly and the dip in the measured IL centered around $1250\,\mathrm{Hz}$ is not seen in the model. The two bubble curtain setup also outperforms the single bubble curtain setup for frequencies exceeding $250\,\mathrm{Hz}$ in line with the measured IL. Comparing Figures \ref{fig:ILSWEEP}a and b, we notice that the IL also depends on the receiver location in the model. In general, the qualitative trend for the variations in the sound pressure levels at the different receivers (see Figure \ref{fig:Alllines}) is captured by the model. 

\begin{figure}[!ht]
\centering
\includegraphics[width=\columnwidth]{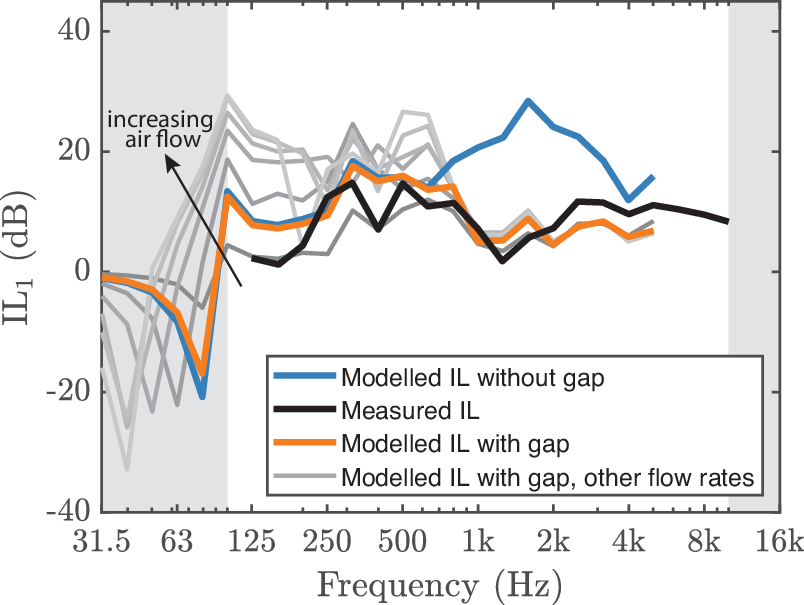}
\caption{(color online) IL of the one bubble curtain configuration, we compare the model with and without a gap in the bubble curtain to the measured IL at an airflow rate of $96\,\mathrm{Lmin^{-1}}$. The grey lines are the results for different air flow rates.} \label{fig:EffectofOpen}
\end{figure}

From Figures \ref{fig:ILSWEEP}a and c it becomes clear that the modelled IL depends on the airflow rate and changes in this quantity not only affect the magnitude but also the frequency dependence of the modelled IL. A similar dependence was not observed in our measurements. To better understand the discrepancy between the measurements and the model we turn to the  version of the model for which we included a gap in the curtain close to the manifold (see section \ref{subsec:Transparency}). The corresponding model result at an airflow rate of $96\,\mathrm{Lmin^{-1}}$ at receiver 4 is compared to the results shown in Figure \ref{fig:ILSWEEP}b for the one bubble curtain configuration in Figure \ref{fig:EffectofOpen}.

It can be seen that including the gap results in a better agreement of the modelled IL with the measured data. Up until $\sim 600\,\mathrm{Hz}$ the influence of the gap is minimal in the model, however for larger frequencies a significant drop in the IL is observed if the gap is included. The frequency of $\approx 600\,\mathrm{Hz}$ at which the gap becomes important does not correspond directly to the length scale of the gap ($1481/0.2\approx 7500\,\mathrm{Hz}$), but this cutoff is found to decrease with increasing gap height (not shown). \citet{wang2023transmission} looked into the effect of a gap more closely. When including the gap, the modelled IL for frequencies exceeding $600\,\mathrm{Hz}$ becomes largely independent of the air flowrate. Including the gap in the model improves the resulting prediction for the measured IL. Even if the present way of account for this effect is rather crude, these results highlight the importance of accounting for openings in the bubble curtain in the modelling.

\begin{figure}
\centering
\includegraphics[width=\columnwidth]{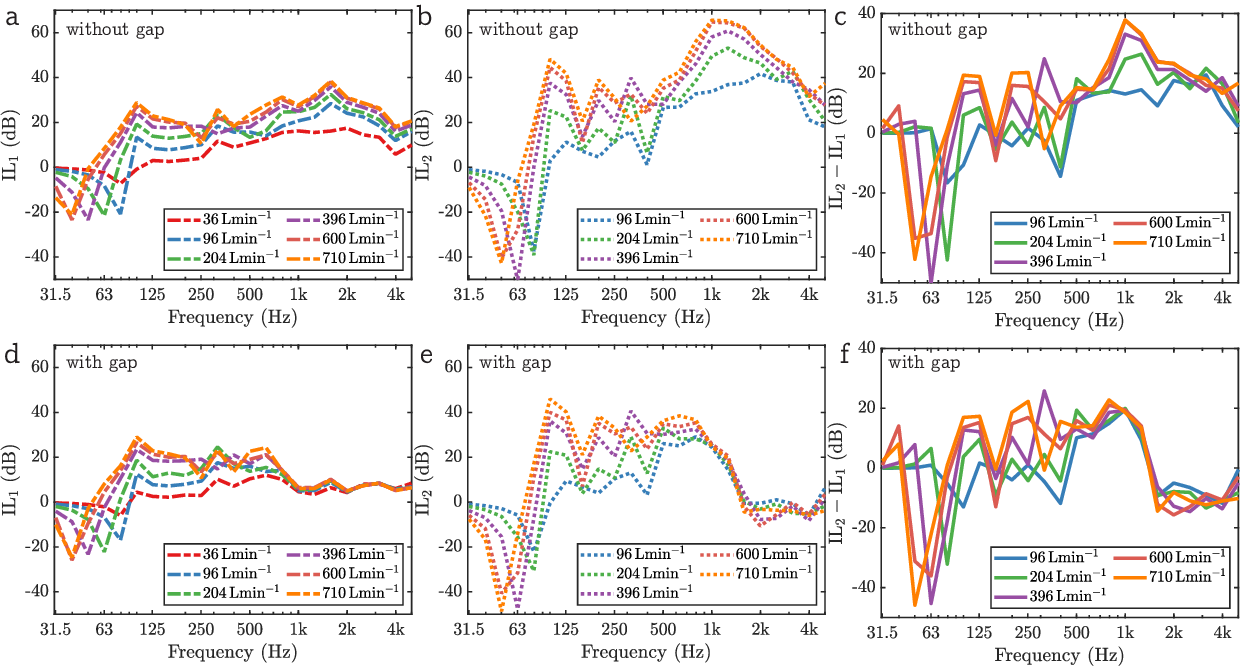}
\caption{(color online) The modelled IL for a) one bubble curtain without gap and b) two bubble curtains without gap c) difference in IL between different configurations without gap d) one bubble curtain with gap e) two bubble curtains with gap f) difference in IL between different configurations with gap} \label{fig:ILModel}
\end{figure}

In the following we will show model results in order to compare the one and two bubble curtain setups from a modelling point of view. Both configurations are tested with and without the gap. The modelled IL without the gap for a single bubble curtain is shown for all flow rates in Figure \ref{fig:ILModel}a. In Figure \ref{fig:ILModel}b the IL is shown for two bubble curtains employed, in Figure \ref{fig:ILModel}c the difference in IL is shown. 

The modelled results show, similar to the measurements, that the two bubble curtain setup increases the IL significantly. Contrary to the measurements, the IL does depend on the airflow rate (as also seen in Figure \ref{fig:ILSWEEP}a and c). For frequencies exceeding $\sim 250\,\mathrm{Hz}$, the model shows an improved IL when using two bubble curtains. The large dip below $125\,\mathrm{Hz}$ in the difference in IL is mainly a consequence of the shifting of the dips between using one and two bubble curtains. \\
Figures \ref{fig:ILModel}d-f show the results for the model with the gap similar to Figures \ref{fig:ILModel}a-c.
For both the one (Figure \ref{fig:ILModel}d) and two (Figure \ref{fig:ILModel}e) bubble curtain configurations the IL becomes flowrate independent for the higher frequencies. However, when employing 2 bubble curtains the IL becomes negative for higher frequencies. Such a behaviour may be expected at lower frequencies due to a possible reduction of the impedance mismatch between the source and the medium. However, the origin of this phenomenon at higher frequencies remains unclear.
Results for the the model including the gap suggest that for higher frequencies inserting two bubble curtains might actually be worse than no bubble curtains since the IL becomes negative, but this is not seen in our experiments (Figure \ref{fig:ILSWEEP}d). For the lower frequencies the effectiveness of splitting up the airflow rate is largely retained even when the gap is included.

The sharp negative peaks around $\sim 80\,\mathrm{Hz}$ in Figures \ref{fig:ILModel}a,b,d,e appear in the IL predicted by the models with or without gap. We could not confirm the presence of such a peak experimentally, since the source does not generate sufficient sound pressure at these low frequencies. 

\subsection{Airgun measurements}
So far we have focused on the results measured by the hydrosounder since the source signal is more repeatable resulting in more reliable measurements (see appendix \ref{app:soundsource}). However, an advantage of the airgun is that it generates more sound, particularly at lower frequencies, and the corresponding results are therefore discussed in the following. 

\begin{figure}[!ht]
\centering
\includegraphics[width=\columnwidth]{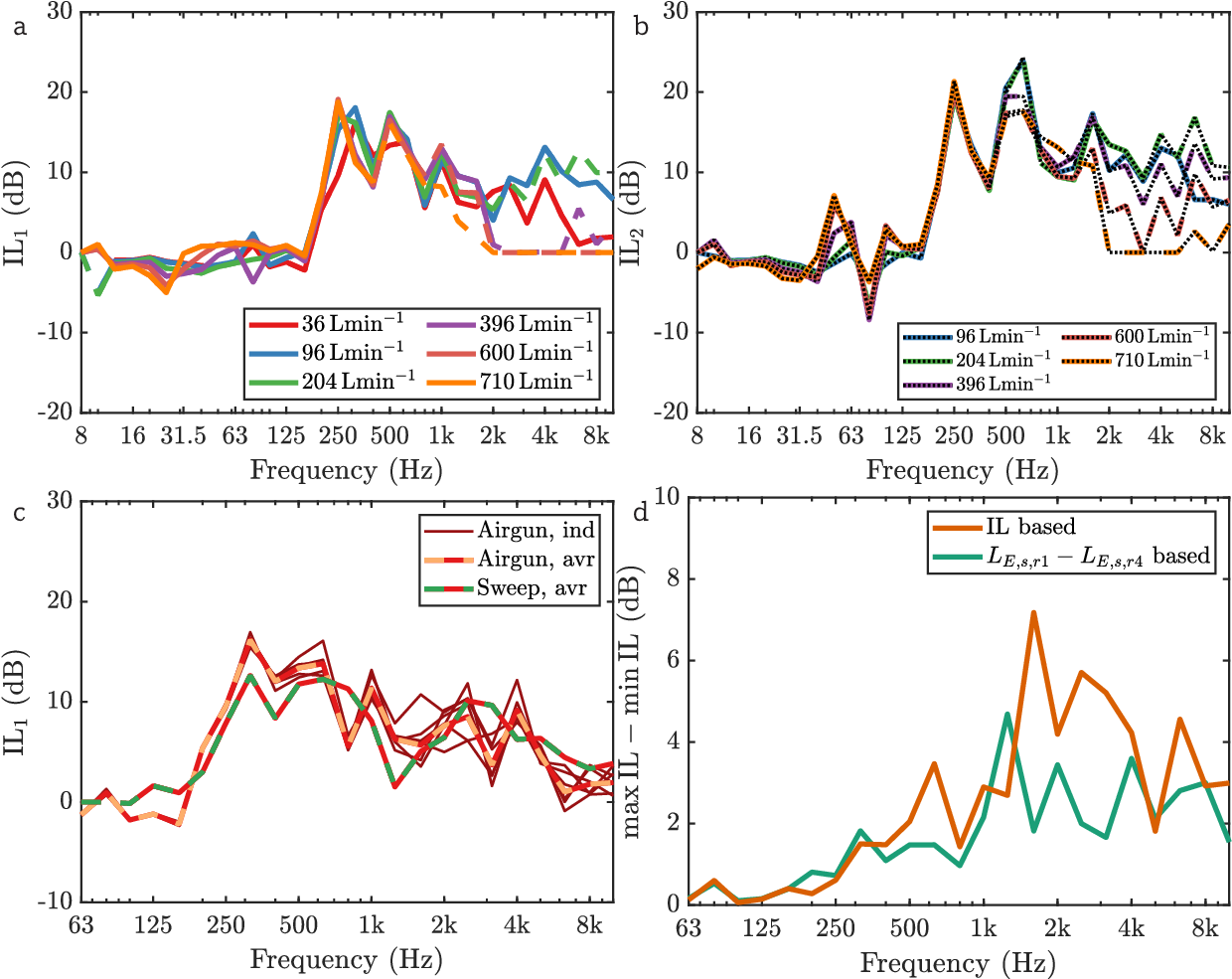}
\caption{(color online) The measured IL by using the airgun for different airflow rates a) for a single bubble curtain and b) for two bubble curtains c) The IL of individual airgun shots and average IL both for the airgun and hydrosounder having a single bubble curtain employed at the lowest airflow rate $36\,\mathrm{Lmin^{-1}}$ d) The maximum variation in IL based on the 6 individual airgun shots} \label{fig:ILgun}
\end{figure}

In Figures \ref{fig:ILgun}a and b we show the IL at the fourth receiver for different airflow rates for one and two bubble curtains respectively. For frequencies exceeding $\sim 500\,\mathrm{Hz}$ the IL derived from the measurements is for more and more flow rates still determined by the noise floor even when using the slightly louder airgun. Nevertheless, the data show that, where measurable, the IL is indeed hardly flow rate dependent. A deviation from the flow rate independent IL can be seen for frequencies exceeding $2\,\mathrm{kHz}$ by comparing $36\,\mathrm{Lmin^{-1}}$ and $96\,\mathrm{Lmin^{-1}}$ in Figure \ref{fig:ILgun}a. 
This could indicate the importance of damping at higher frequencies, but this seems unlikely as it is not consistent with our previous observations and is not supported by the results from the model with and without a gap. An alternative, more likely, explanation is that the bubble curtain was not fully developed at this low air flowrate, underlining the relevance of openings in the curtain discussed in section \ref{sec:res_model}. Comparing Figures \ref{fig:ILgun}a and b paints a similar picture as to the measurements with the hydrosounder in that the overall IL increases when splitting up the airflow over two manifolds.

In Figure \ref{fig:ILgun}c the IL is plotted for individual shots of the airgun alongside the average IL with the airgun and the hydrosounder for the lowest airflow rate of a single bubble curtain ($36\,\mathrm{Lmin^{-1}}$). Quantifying these variations is important since they cannot be predicted by models using average descriptions of the bubble curtain. There are two main possible sources of variation in the presented IL: The (non) repeatability of the source and the variations in the bubble curtain composition. It is known that the source level spectrum becomes less repeatable for subsequent shots for frequencies exceeding $630\,\mathrm{Hz}$ for this particular airgun (see appendix \ref{app:soundsource}). To distinguish the two sources of variation we show two properties. The spread in the IL will be defined as $\max{\mathrm{IL}}-\min{\mathrm{IL}}$ for the individual airgun shots and is shown in Figure \ref{fig:ILgun}d, which we call the IL based result. The variations can be as much as 7 dB. To estimate the contribution of the variation in the sound source to the spread in the IL, the spread in the difference between the SEL at receiver 1 and 4 is plotted. Since both of these levels are measured using the same source signal, the spread is therefore more likely to be mainly due to time dependent properties of the bubble curtain(s) (in appendix \ref{app:ILvar} we discuss the relevance of this property in more detail). Even though the variations determined in this way are generally lower, they can still exceed 4 dB, suggesting changes in the composition  of the curtain play a significant role in the observed spread in IL.

\section{Discussion\label{sec:5}}
In this section we will discuss the results with regard to our objective of investigating the effect of a two bubble curtain configuration. However, we can not interpret these results without considering the limitations resulting from our experimental setup.

\subsection{Remarks on the experimental setup}
Some of the advantages of measuring inside a basin are the repeatability, the known geometry and the low noise floor. In addition, hydrodynamic measurements can be conducted alongside acoustic measurements in a fresh water environment for which our hydrodynamic model \citep{beelen2024planar} was validated previously. However, measuring the acoustic performance of bubble curtains in a basin puts limitations on the direct applicability of the results and conclusions to situations that are closer to the use-case of interest, being impact pile driving in ocean waveguides.\\
\\
To be more precise, the bubble curtain is placed inside a basin and its performance cannot be seen independent of this environment. Placing an omni-directional sound source, such as the ones used here (for the frequency range of interest), in an enclosed basin leads to an interference pattern due to interactions of the sound field directly radiated by the source and its reflections on the side walls and on the water surface. As a result, the amplitude of the sound field strongly depends on the measurement location within the interference pattern. This becomes problematic when interpreting the IL since the presence of a bubble curtain alters the interference pattern. A location at which constructive interference happens at a specific frequency without the curtain may exhibit destructive interference when the curtain is present, and vice versa. The IL, therefore, does not only represent the change due to inserting a curtain for the direct path, but is also influenced by the change in contributions of paths resulting from interaction with the basin walls and the water surface. The influence of secondary paths can be suppressed by adding an array of hydrophones behind the curtain/curtains in future measurements instead of a single hydrophone. Such a setup could potentially give insight in the relative contribution of the direct path vs other paths.

Having observed the different IL curves based on measurement and model data, we notice that they are very spiky, which could partially be caused by the interference between different sound paths. We believe that despite the influence of the basin environment, the trends in the IL as measured in the basin are qualitatively relevant for comparative analysis between different cases and real life bubble curtain applications. For instance, while the difference in IL curves for one and two bubble curtains might be very different for specific frequencies due to changes in interference patterns, the overall observed trend is that the IL is higher for a two bubble curtain setup.

Besides the effect of the influence of the environment on the IL it was also observed that for most airflow rates both sources used during the measurement are incapable of generating a signal at the other side of the curtain that exceeds the noise generated by the bubble curtains. This implies that an accurate assessment of the IL over the complete frequency range cannot be made for many of the investigated cases.

\subsection{Effect of one or two bubble curtains}

With the experiments we set out to test the effect of different bubble curtain configurations on the insertion loss. As can be seen in Figure \ref{fig:ILSWEEP}d the effect of splitting the total airflow rate over two distinct bubble curtains is considerable. Splitting up the airflow into two separate bubble curtains generally increases the performance of the supplied air within the measurable range. The effect of the split, however, does not manifest itself as a constant increase in IL that is observed for all frequencies. Moreover, there are decidecade bands, for which the increase is high with neighboring bands for which the increase is almost zero or for which a decrease is observed. A conspicuous deviation to the increased performance of the bubble curtain when the supplied air flow rate is increased occurs around $250-400\,\mathrm{Hz}$. This corresponds roughly to $c/\Delta_{bc}\approx 250\,\mathrm{Hz}$, which seems to be an eigenfrequency of the system with inter curtain distance $\Delta_{bc}$. The main reason as to why this split up of the total airflow rate is effective is due to the non-linear relation between the speed of sound and the local void fraction, as mentioned in the introduction. For a typical void fraction along the centerline of the curtain of $\sim 1\%$ the local speed of sound, according to \citet{wood1956textbook}, is $c_m\approx 120\,\mathrm{ms^{-1}}$. Halving the void fraction to $0.5\%$ results in a speed of sound of $c_m\approx 170\,\mathrm{ms^{-1}}$. Both of which remain low as compared to the speed of sound in water of $c=1481\,\mathrm{ms^{-1}}$, and therefore maintain a large impedance mismatch. Primarily, the positive effect of adding a second bubble curtain can therefore be explained by the maintained impedance mismatch for each bubble curtain even as the airflow rate per curtain is reduced. This effect is mainly important for the lower frequencies ($<1000\,\mathrm{Hz}$), for which reflection is the dominant mechanism for sound reduction. The independence of the IL from the airflow rate, as seen in Figures \ref{fig:ILgun}a and b, is another indication of the probable importance of this mechanism. The observed flow rate independence of the IL, however, must be interpreted with caution, since the experimental setup may influence this conclusion. Nevertheless, \citet[e.g.][]{rustemeier2012underwater} reach similar conclusions and report only a small dependence of the IL on the airflow rate, compared to the large dependence on the configuration.  
For increasing frequency, approaching and exceeding the eigenfrequencies of the present bubbles, the split up of the airflow rate is expected to be less effective as damping becomes more relevant compared to reflection. For very low frequencies ($f\ll c/\Delta_{bc}$) the two bubble curtains might act as one. The latter point as well as the observed lower performance around the eigenfrequency of the system demonstrate that the distance between the two manifolds influences the performance of the system.

The model predicts an increased effectiveness of splitting up the airflow rate into two bubble curtains just as the measurements show. The magnitude of the improvement, and the absolute values of the IL are however overpredicted by the model.
The very significant increase in IL when using two manifolds is interesting for real life applications of bubble curtains. The results show that compressed air can be used more effectively when it is split up in two distinct bubble curtains. Splitting the air over even more manifolds, while keeping the inter bubble curtain distance the same could prove to be even more effective use of the compressed air. 

\subsection{Model vs reality}
\label{subsec:disc_modelvsreal}

In Figure \ref{fig:ILgun}c we showed the large variation in the IL for individual shots with the airgun. This spread originates from both the variation in the source sound levels and the variability of the effectiveness of the bubble curtain, as shown in Figure \ref{fig:ILgun}d. We used a time interval of 5 seconds between different instances of the airgun, such that effectively all the air inside the bubble curtain has been replaced (3.6 m deep with a rising velocity of $\sim 1\,\mathrm{ms^{-1}}$). Each sound pulse thus "sees" a new and different bubble curtain. Some of the effects that change the bubble curtain in time are the wandering, the cloud like structures with locally different bubble densities, and the deformation along the length of the manifold (\citep[ e.g.][]{beelen2023situ}). All of these temporal variations contribute to the spread in the IL. The cloud like structure in particular, has a known effect on the transmission of sound waves \citep[e.g.][]{d1988acoustical,kozhevnikova1992experimental,omta1987oscillations} and changes in the structure can thus change the acoustic properties of the bubble curtain. Further variations in the structure of the curtain can have an effect on the sound transmission if the typical size of these structures is large as compared to the wavelength of a sound wave. Equivalent fluid models using an average description of the bubble curtain profiles, will not be able to describe the fluctuations in the IL over time. The variations in the IL show that the actual distribution of the air in the bubble curtain is important. Furthermore, taking the average of the acoustic properties based on instantaneous void fraction distributions for many instances of a bubble curtain in time does generally not equal the acoustic properties based on the average void fraction distributions for the same instances of the bubble curtain. The bubble curtain representation using the average void fraction distribution could, therefore, lead to inaccurate acoustic predictions. It is currently unknown how variations in the instantaneous distributions describing the bubble curtain affect the acoustic performance of the bubble curtain as compared to the performance of a theoretical bubble curtain with the average void fraction distribution. For instances of the bubble curtain featuring a larger number of holes, the performance will likely deteriorate. On the other hand, if at the same time bubble clouds respond to a more desirable frequency it might improve the relevant performance. 

Another shortcoming of the acoustic models, as they are currently used, is related to equivalent fluid modeling. Experiments aimed at solely measuring the response of a typical bubbly mixture for bubble curtains (gas fraction on the order of $\sim 1\%$ and bubble radii in the order of of $\sim 2\,\mathrm{mm}$) to sound have not been carried out. \citet{commander1989linear} mention that their model is not applicable in the ranges that it is currently used for in bubble curtains. We would like to shortly discuss a concept for measuring the isolated bubble plume response to sound, since the lack of validation data remains problematic. If a hydrophone is placed inside a bubble plume (i.e. inside the bubbly mixture) while keeping the source outside this plume likely all sound sources other than the sound generated by the hydrosounder or airgun are negligible. Since the hydrophone is inside the bubble plume it is as much shielded from the background noise, the noise generated by the bubble generation and the indirect sound paths as it is from the direct sound. Since the direct sound is significantly louder than the other sounds they can be neglected. This is assuming the sound generated by the bubble plume stems from the manifold and the water surface, not from the rising of the bubbles itself. By measuring inside the bubble plume one could get more insight on the isolated response of the bubble curtain to the generated sound. Moreover it can be incorporated in experiments such as the ones explained in this paper by moving one of the hydrophones to within the bubbly mixture.

\subsection{Holes in the bubble curtain}
From the theoretical considerations and from the model results it seems that voids in the bubble distribution are likely an important factor for the characterization of bubble curtains. Holes in the bubble curtain can be considered as sources of sound leakage. Close to the manifold the presence of holes is most obvious. Representing these holes as a region with the speed of sound of water in our 2.5D FE model brings the modelled results closer to the measured results (see Figure \ref{fig:EffectofOpen}). This simulations with the gap shows that the sound transmitted by the relatively small gap can be dominant over the sound transmitted by the much larger bubble curtain. 
 A more thorough study including experiments and modelling is needed to investigate the role of openings in the curtain in more detail.\\

In this context, it is also noteworthy to consider the results of \citet{rustemeier2012underwater}, who reported a significantly improved performance for a bubble curtain emanating from a membrane instead of individual nozzles. These authors named the reduced bubble size as a potential reason for the increased insertion loss when using the membrane. However, this effect could not explain the full performance improvement observed in the experiments. It therefore appears plausible and likely, that the transparency, which is greatly reduced when using the membrane, also is a relevant factor here.  
\\
The cloud-like structure in a developed bubble curtain could lead to local holes in the bubble curtain, temporarily increasing the transparency. This could partially explain the variability in the performance of the bubble curtain between different shots of the air gun (see Figure \ref{fig:ILgun}d). \\
Finally, even if the gap is included in the model, the model results retain a flow rate dependence at lower frequencies. This is not seen in the experimental data. Interestingly, the experimentally observed flowrate dependence of the IL between the flowrates $36\,\mathrm{Lmin^{-1}}$ and $96\,\mathrm{Lmin^{-1}}$ (see in Figure \ref{fig:ILgun}a) is actually not seen for the model including the gap. Note however that the size of the gap in the model is not dependent on flow rate. It therefore appears possible that larger effective openings where present at those lowest flow rates (e.g. not all nozzles operating or reduced lateral mixing), reducing the IL for these cases.

\section{\label{sec:6}Conclusion and Outlook}

In this paper we have investigated the impact of the configuration of bubble curtain(s) on the insertion loss based on measurements and on modelling. The experiments show that splitting the airflow rate into two distinct bubble curtains significantly increases the IL for frequencies exceeding $100-500\,\mathrm{Hz}$, which is relevant for pile driving. Similarly the modelling indicates a significant improvement when splitting up the airflow rate. The distance between the bubble curtains will influence the frequency range in which splitting up the air flow rate over two manifolds is effective. We attribute the improvement in IL when splitting up the airflow rate to the introduction of a secondary reflective boundary without compromising the effectiveness of the individual curtains too much, since the speed of sound increases only marginally due to the reduction in flow rate per curtain. This study shows there is room for improvement in the performance of bubble curtains without increasing the net air flow rate when considering the configuration as part of the optimization strategy. These benefits need, of course, to be weighed against the cost of installing an additional bubble curtain. For single bubble curtains our results show that using lower air flow rates may have only a limited impact on the performance.

Besides these findings regarding the performance of the considered air curtain configurations, a number of other insights are noteworthy. We have seen that the IL does not depend on the flow rate in our measurements, whereas it does when considering the model results using the the average void fraction to compute the effective speed of sound. We did obtain a flow rate independent result for higher frequencies when we consider the presence of an opening (e.g. close to the nozzles) in the modelled bubble curtain. This finding requires further investigation. The effectiveness of the gap indicates it's importance, however, we want to stress that this finding does not mean that this was the missing part in the further correct modelling. The effectiveness of the gap can also disguise the fact that the equivalent fluid models currently employed are not suitable for the considered conditions (i.e. void fraction range and frequencies close to or at bubble resonance).\\
Another issue that deserves further investigation is the effect of using the average void fraction distribution to calculate the acoustic properties instead of using multiple instances of the instantaneous void fraction distribution to calculate the acoustic response and average over that. To accurately capture the details of the bubble curtain structure that exist at each instance and its effect on the acoustic performance, 3D simulations would be required. However, also 2D simulations including variations in the bubble curtain in the direction perpendicular to the manifold, could already provide important insight in its effect.

For many of the higher air flow rates a louder sound source would be required to increase the frequency range in which the measured signals are above the noise floor and the IL can correctly be assessed. At the same time, the indirect paths of the sound should be limited as much as possible so that experimental results in basins can be translated more easily to real life applications. The latter might be impossible in a basin such as the one used for our experiments. To isolate the effect of a bubble curtain, from its surrounding we suggested in section \ref{subsec:disc_modelvsreal} to place a hydrophone inside a bubble curtain. These experiments could function as a benchmark case for acoustic modelling of bubble plumes. Obviously, these acoustic measurements should be accompanied by hydrodynamic measurements of the void fraction and bubble size distribution. 

Further research in applying two bubble curtains, using the same airflow rate, for full scale applications should also be carried out. For example our finding implies that halving the airflow rate per curtain benefits the overall IL, but this only works if the bubble curtains remain continuous. The potential for up-scaling this approach should therefore be carefully investigated.

\begin{acknowledgments}
We gratefully acknowledge the support by TNO and MARIN and in particular Frans Staats, Martijn van Rijsbergen and Heerko Koops during the measurements. We would also like to thank Gert-Wim Bruggert and Thomas Zijlstra for their help in preparing the experimental setup. 
\end{acknowledgments}

\section*{Author Declarations}
\subsection*{Conflict of Interest}
The authors declare no competing conflicts of interest.
\subsection*{Data Availability}
The data used in this study are available from the corresponding author upon reasonable request.

\appendix
\section{Logarithmic sweeps J11} \label{app:appendixAch3}

\begin{figure}[!ht]
\centering
\includegraphics[width=\columnwidth]{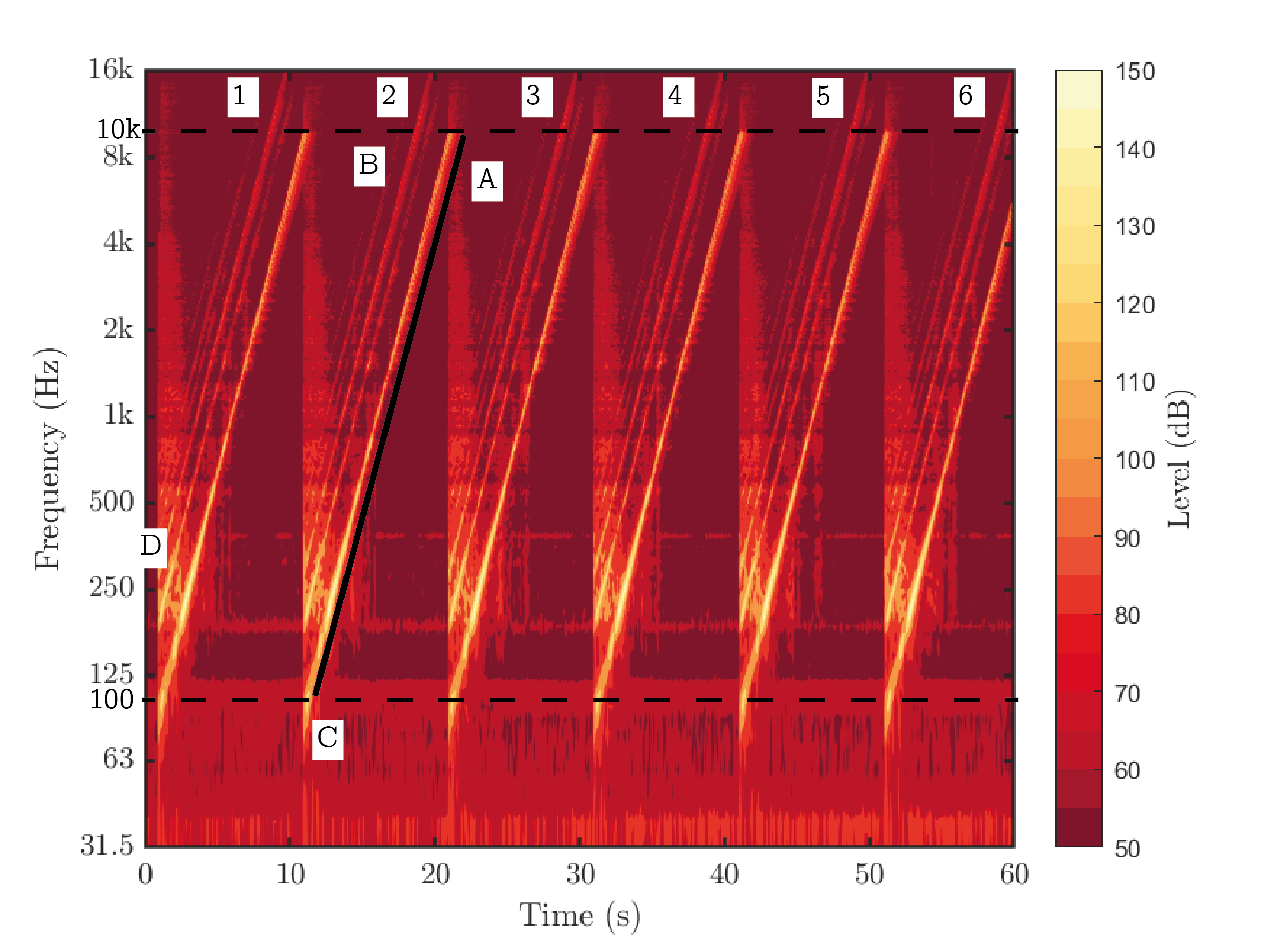}
\caption{(color online) Decidecade band spectogram of the sound pressure sweeps from the J11 hydrosounder, recorded by receiver 2 without a bubble curtain active} \label{fig:logsweeps}
\end{figure}

In Figure \ref{fig:logsweeps} we show the 6 logarithmic sweeps generated by the J11 hydrosounder per measurement. The frequency sweeps are linear against the logarithmic $y$-axis as shown by the line (A). Along with the main signal the source also generates higher harmonics as is evident from the parallel lines (B). The J11 also unintentionally generates lower frequencies (C). These higher and lower frequencies can sometimes be useful for analysis outside of the intended range although they do lack power. Finally we see a slight lack in the synchronization (D) such that the sixth sweep is not entirely captured. We therefore use the first 5 sweeps.

\section{Model implementation in Comsol} \label{app:appendixB}

In this appendix we discuss details regarding the implementation of the model in Comsol.\\

\begin{figure}[!ht]
\centering
\includegraphics[width=\columnwidth]{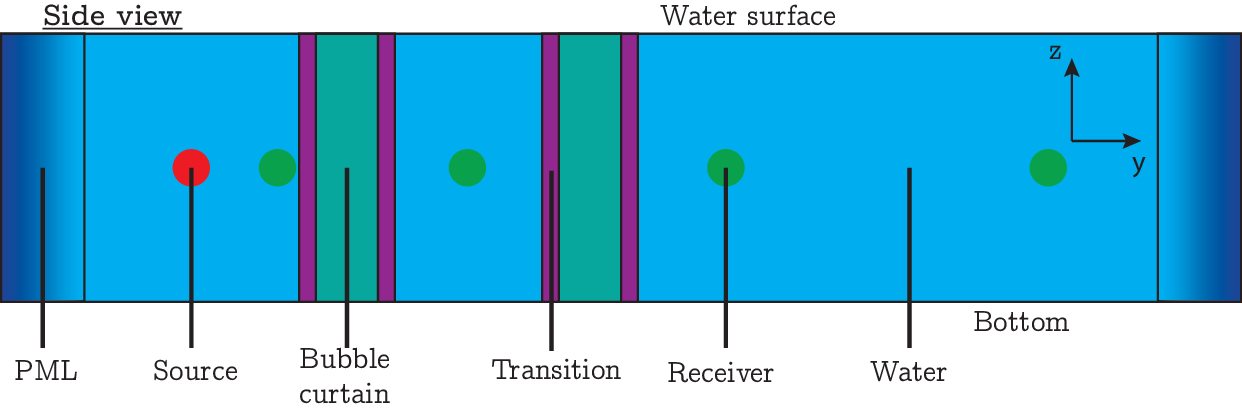}
\caption{(color online) Illustration of the modelled domain and the locations of the source and receivers} \label{fig:Model}
\end{figure}

\noindent\emph{Domain}\\
An overview of the modelled domain is given in Figure \ref{fig:Model}. It consists of a pure water domain (blue in Figure \ref{fig:Model}), two domains where the bubble curtains are positioned (cyan in Figure \ref{fig:Model}) with accompanying transition domains (purple in Figure \ref{fig:Model}) and on both ends a so-called perfectly matched layer (PML). We will discuss the implementation of each domain in the following subsections. 

The location of the source is shown in red. The location of the source and receivers correspond to the locations of the source and receivers in the experiments. The four receivers are shown by the green dots. The boundaries are defined by the PML's (explained below) and by the bottom and water surface.\\

\noindent\emph{Bubble curtain}\\
The complex valued equivalent sound speed of the bubble curtain $\langle c_m\rangle(y,z)$ is saved into a look up table, which is interpolated to the Guass integration points of the local mesh in the domain assigned for the bubble curtain.\\

\noindent\emph{Perfectly matched layer}\\
The perfectly matched layer (PML) was first introduced by \citet{qi1998evaluation} for acoustic problems. The PML is a non-physical material which is used as a boundary to a computational domain. This boundary compactly and efficiently describes the damping of unbounded media. In other words the PML represents a long open domain in which the acoustic pressure is dampened. In Comsol a PML is readily implemented (see e.g. \citet{zampolli2008improved}) and has been applied for the currently presented model. As the name suggests the perfectly matched layer retains the same properties as the material it is supposed to represent, this entails that, although there is a lot of damping, no reflections will result from this implementation.\\

\noindent\emph{Mesh}\\
The mesh will be auto-generated by Comsol given some manual restrictions on the mesh size. For an accurate description of the solution, using quadratic elements, the mesh size ($s_{\mathrm{mesh}}$) needs to be smaller than a quarter of the smallest wavelength considered

\begin{equation}
    s_{\mathrm{mesh}}\leq \frac{1}{4}\frac{c_m}{ f_{\mathrm{max}}}=\frac{1}{4}\lambda_{\mathrm{min}}.
    \label{eq:meshsizecond}
\end{equation}

\noindent
For the cases at hand, $f_{\mathrm{max}}=5495\,\mathrm{Hz}$ is the maximum frequency considered for the solution and $\lambda_{\mathrm{min}}$ is the accompanying minimum wavelength. In the water and PML domains this condition has been implemented by $1/8 \lambda_{\mathrm{min}}<s_{\mathrm{mesh}}<1/4 \lambda_{\mathrm{min}}$, with $\lambda_{\mathrm{min}}\approx 0.27\,\mathrm{m}$. The mesh in the water domain is a rectangular mesh fulfilling the aforementioned condition for the smallest side of the rectangle. In practice the mesh size is as large as possible while fitting well within the geometric constrictions of the domain, such that it is closer to $1/4\lambda_{\mathrm{min}}$. \\
In the bubble curtain domain we also chose to employ a constant mesh size, for that the minimum sound speed in the bubble curtain has to be set beforehand. However the sound speed can locally be as low as $\langle c_{m,\mathrm{min}}\rangle \approx 20\,\mathrm{ms^{-1}}$, resulting in a very dense mesh. It must be noted that these very low sound speeds are only observed in very narrow parts of the center of the bubble curtain for specific frequencies and curtain configurations. To alleviate the condition on the mesh size (see Eq. \ref{eq:meshsizecond}) we experimentally tested with different effective minimum sound speeds, to find $\langle c_{m,\mathrm{min}} \rangle=75\,\mathrm{ms^{-1}}$ to still faithfully describe the solution. The shortest wavelength considered and thus determining the mesh size is $\lambda_{\mathrm{min}}\approx 13.6\,\mathrm{mm}$ and is more than sufficient for the vast majority of the bubble curtain domain. A further criterion on the mesh size is that it should be able to describe the local changes in the speed of sound accurately which was experimentally found to be the case using the aforementioned mesh size. The mesh in the bubble curtain domain is a triangular mesh allowing for more freedom in the mesh generation, although not relevant for the present study this is implemented to allow for the option of a non-rectangular shape of the bubble curtain domain.\\
Finally the transition region connects both meshes by free triangulation. This allows for a smooth transition between both meshes, with mesh sizes between the two aforementioned mesh sizes.\\

\noindent\emph{Propagation modes}\\
As mentioned before a 2.5D model does not simulate the 3rd dimension using FE simulations but assumes a solution to the Helmholtz equation fitting the boundary conditions. For Cartesian coordinates the solution to the 1D Helmholtz equation satisfying acoustically hard boundary conditions takes the form of cosine functions. The excitation, which is a point excitation in the middle of the basin is expanded into this set of cosine functions to obtain the contribution of each mode. The solution will only include the symmetric cosines, since these can be actuated by the source and fit within the symmetric domain. In the FE calculation, which is needed to obtain the $y$-$z$ component of the solution, each mode uses an augmented wave number

\begin{equation}
    k_a^2=k_{y,z}^2=k_{x,y,z}^2-k_x^2.
    \label{eq:augmentedwavenumb}
\end{equation}

\noindent
The augmented wave number, $k_a$, or $k_{y,z}$, is effectively the projection of the 3D wave number on the 2D plane along the line the source is located ($x=x_s=0$) \citep{cao19972,novais20052}. $k_x$ is the wave number of the solution in the $x$-direction and $k_{x,y,z}$ is the local wave number of the fluid or mixture ($c/\omega$ or $c_{m}/\omega$ respectively with $\omega=2\pi f$). Since the source is placed in the center of the domain only the even contributions of the cosines and a constant solution need to be taken into account. We have taken the first five even modes into account, leading to minimal truncation errors while maintaining a manageable computational effort. Eq. \ref{eq:augmentedwavenumb} indicates that for large $k_x$ relative to the real part of $k_{x,y,z}$ the imaginary part of the augmented wave number, $k_a$, becomes dominant yielding evanescent waves. The higher modes are thus generally dampened quickly and hardly contribute to the overall solution especially at larger distances from the source. To retrieve the approximated 3D solution the 2D solution, calculated using the augmented wave number, needs to be multiplied with the associated mode shape in the $x$-direction and then summed. However, since the receivers (the hydrophones) are, just like the source, placed in the symmetry plane we can simply sum the 2D solutions of the different modes to find the solution in this plane.\\

\noindent\emph{Source}\\
Since we represent the result in terms of the insertion loss the source representation is not a critical factor in the model i.e. we are only interested in the sound reduction by the bubble curtain(s). In reality the directionality of the source could, however, influence the measured IL. The source is implemented as a point source, since the source in the experiments is roughly omnidirectional. The coordinates of the source are $x_s$, $y_s$ and $z_s$ and correspond to the locations of the source in the experiments.\\

\noindent\emph{Receivers}\\
The solution is interpolated to the receiver locations to correspond to the locations of the hydrophones in the experiments. For these receiver points the IL was calculated according to Eq. \ref{eq:ILdefSPL}. The coordinates of the receivers are $x_{r,i}$, $y_{r,i}$ and $z_{r,i}$ with $i=[1, 2, 3, 4]$.\\

\noindent\emph{Boundary conditions}\\
At the water surface the acoustic pressure vanishes, known as the pressure release condition or as the \lq sound soft boundary\rq in Comsol. At the bottom of the tank and at the end of the PML's the normal component of the velocity vanishes, known as the \lq sound hard boundary\rq\, in Comsol. The latter boundary condition approximates the walls of the basin, however, the impedance mismatch between the water and the concrete of and the soil against the basin walls is limited. This boundary condition thus omits the interaction between the water and the walls and soil and the secondary path this creates in practice.

\section{The effective speed of sound and transparency}
\label{app:SpeedofsoundAVG}
In this appendix we explore the effect of holes in a bubble curtain. Currently these structures are generally not taken into account in the equivalent fluid modelling of bubble curtains. The equivalent fluid model usually assumes a homogeneous bubble distribution after which the equivalent speed of sound is calculated, which in turn is input to the 2.5D simulation. In this appendix we will discuss, using the theory that is the basis to the equivalent fluid modelling, why disregarding holes in the bubble curtain could be problematic and what the best (or least bad) option is for calculating the effective speed of sound in the presence of these holes is.

\begin{figure}[!ht]
\centering
\includegraphics[width=\columnwidth]{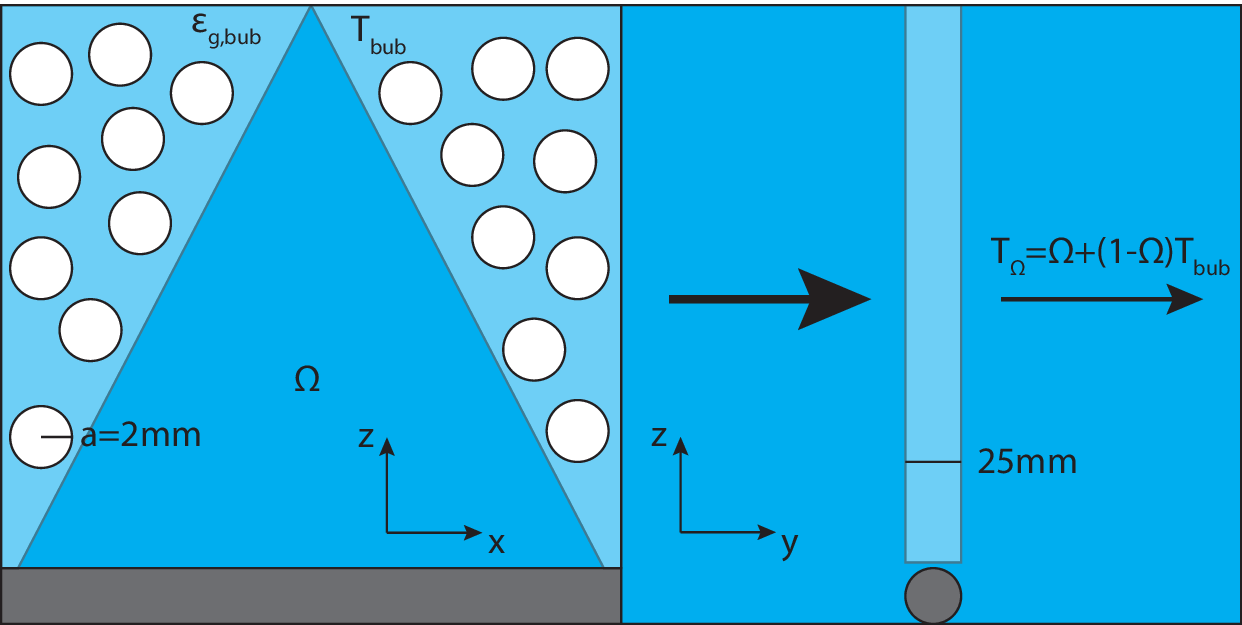}
\caption{(color online) Schematic overview of the structures close to the manifold.} \label{fig:Transcoeffsketch}
\end{figure}

Bubble curtains are typically generated from an array of individual round plumes, which eventually merge to form a continuous quasi-2D bubble plume. This leads to an inhomogeneous distribution of the void fraction in the region close to the nozzles, which is also reflected in the hydrodynamic model used to calculate the void fraction and bubble size distributions. This variation in the void fraction along the $x$-direction, in turn, leads to a variation of the speed of sound in the same direction. The reduction to an effective speed of sound required for the 2D model can then be achieved in multiple ways, of which we take the two most common, namely

\begin{equation}
c_{eff}=\langle c_m\rangle_x (y,z)
\label{eq:avgspeedofsound}
\end{equation}
\noindent
or,
\begin{equation}
c_{eff}= c_m (\langle \epsilon_g \rangle_x,y,z),
\label{eq:avgspeedofsoundVF}
\end{equation}
\noindent
where $\langle\cdots\rangle_x=\frac{2}{\Delta x_n}\int_0^{\Delta x_n/2}\cdots dx$ denotes the spatial average in the $x$-direction. That is,  we can either calculate the speed of sound first and then average (Eq. \ref{eq:avgspeedofsound}) or determine the average void fraction first and then calculate the speed of sound (Eq. \ref{eq:avgspeedofsoundVF}). The described order of averaging is important due to the non-linear relationship between the void fraction and the speed of sound, currently the latter (Eq. \ref{eq:avgspeedofsoundVF} is mostly used). The 2D model aims at representing the actual 3D situation, implying that the effective speed of sound needs to result in the correct transmission of the sound pressure through the bubble curtain(s). We will evaluate the choice for the effective speed of sound based on a simplified theoretical situation sketched in Figure \ref{fig:Transcoeffsketch}, which approximates the void fraction distribution close to the nozzles.

\begin{figure}[!ht]
\centering
\includegraphics[width=\columnwidth]{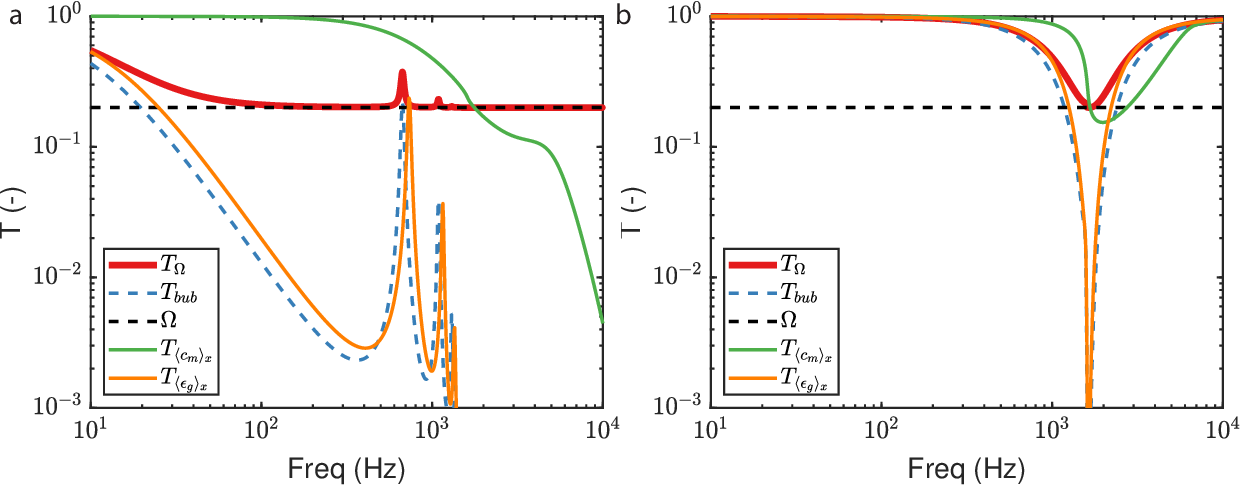}
\caption{(color online) Transmission coefficient based on different approaches of calculating the speed of sound. a) $\epsilon_{g,bub}=0.1$ b) $\epsilon_{g,bub}=0.001$.} \label{fig:Transcoeff}
\end{figure}

For the proposed simplified analysis, we assume the bubble curtain to consist of a homogeneous bubbly regime and a pure water regime. In this way we can compare the transmission resulting from our choice of the speed of sound to the theoretical result. The transparency factor $\Omega$ denotes the fraction of the cross-section in the $xz$-plane which does not contain bubbles. We assume that there is no transmission loss in the region without bubbles, proportional to $\Omega$, such that the power based transmission coefficient $T= 1$ in this region. For the remaining surface, proportional to $1-\Omega$, we adopt the transmission coefficient $T_{bub}$ given by \citet{commander1989linear} for the theoretical reference case. Note that this is an approximation since \citet[p.736]{commander1989linear} corresponds to the transmission coefficient of a homogeneously filled bubbly screen. 
The total theoretical transmission coefficient, $T_{\Omega}$, is then given by the weighted average

\begin{equation}
    T_{\Omega}=\Omega+(1-\Omega)T_{bub},
    \label{eq:EstimateT}
\end{equation}
which results in $T_{\Omega} \to \Omega $ in the limit $T_{bub} \to 0$. We now compare the resulting transmission coefficient of Eq. \ref{eq:EstimateT} with two alternative estimates, viz. a transmission coefficient based on a space averaged sound speed ($T_{\langle c_m\rangle_x}$) according to Eq. \ref{eq:avgspeedofsound} and a transmission coefficient based on an averaged void fraction ($T_{\langle\epsilon_g\rangle_x}$) according to Eq. \ref{eq:avgspeedofsoundVF}. In Figure \ref{fig:Transcoeff} the resulting transmission coefficients are shown.

We use a relatively high void fraction in Figure \ref{fig:Transcoeff}a in the bubble regime of $\epsilon_{g,bub}=0.1$ due to the fact that the structures resulting from the hydrodynamical model occur only close to the manifold where the void fraction is high. In practice however, holes can also occur as a result of the structures due to the cloud like behaviour of the bubbles or as a result of currents at sea. Therefore we also show a low void fraction in Figure \ref{fig:Transcoeff}b. In this way we can roughly estimate the impact of holes higher up in the bubble curtain. We use a homogeneous bubble size distribution with a bubble radius of $a=2\,\mathrm{mm}$, a transparency factor of $\Omega=0.2$ and a thickness of $25\,\mathrm{mm}$ (based on $1/4$ of the inter nozzle distance of the manifolds used in our experiments). \\
In particular in Figure \ref{fig:Transcoeff}a, but also in Figure \ref{fig:Transcoeff}b it can be seen that both transmission coefficients ($T_{\langle c_m\rangle_x}$ and $T_{\langle\epsilon_g\rangle_x}$) do not agree with the estimate in Eq. \ref{eq:EstimateT}. However, unlike $T_{\langle c_m\rangle_x}$,  $T_{\langle\epsilon_g\rangle_x}$ is largely consistent with the behaviour of a homogeneous bubble curtain ($T_{bub}$), and also approximates $T_{\Omega}$ reasonably well at the lower void fraction. The better (or least bad) option is thus to first average the void fraction and then calculate the speed of sound. An additional benefit of this approach is that it will likely be able to represent the so-called wall effect more closely. The wall effect was first discussed by \citep{moens1877over,korteweg1878ueber}. They found that the pressure pulse in blood vessels did not move with the higher speed of sound in the blood but rather with a lower speed of sound depending on the properties of the blood vessel (the wall). In this analogy the bubble plume acts as the blood vessel and the pure water as the blood. The speed of sound through the holes should then be lower than that of water. This effect would thus favor the void fraction averaged result, the openings, however, can not be considered narrow everywhere such that the importance of this effect could be minimal.

A further observation in relation to Figure \ref{fig:Transcoeff} is that potential positive effects through increasing the void fraction or changing the bubble size distribution are largely cancelled out by the effect of the holes in the bubble curtain. The minimal transmission coefficient is namely limited by $\Omega$. This is similar to a door that is ajar, where the total transmission is determined by the opening such that improving the structure has no effect. The door can be seen as the bubble curtain with the transmission coefficient $T_{bub}$ and the fact that it is ajar as the holes, represented by $\Omega$. In that case the fact that it is ajar sets a lower limit to the relevance of the effectiveness of the door. Following this argumentation leads to a flow rate independent IL if the holes are important. However, the environment of the hole (the impedance of the bubble curtain surrounding it) affects the IL. See also the discussion of the 'slit on the ground' in \citet{wang2023transmission}.\\
One possible way of dealing with the inability to find a proper effective speed of sound can be to translate the 3D holes to a 2D gap. In other words replace part of the bubble curtain by pure water, similar to \citet{peng2021study}. However translating the 3D holes to a 2D gap is non-trivial for example due to the before mentioned wall effects. Also the interaction between the holes and the frequency of the incident sound wave is largely unknown.

\section{Background noise} \label{app:backgroundnoise}
\begin{figure}[!ht]
\centering
\includegraphics[width=\columnwidth]{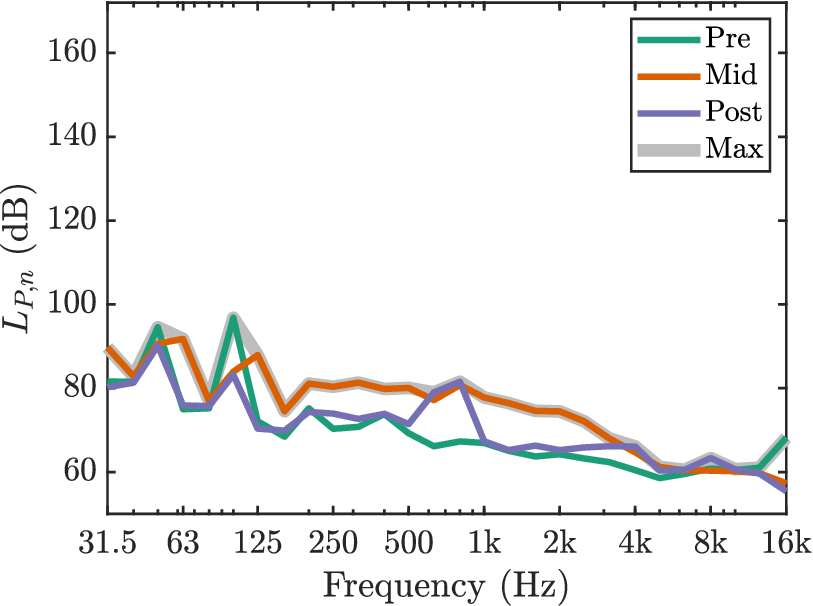}
\caption{(color online) Background noise levels measured at the fourth receiver location at three different time instances} \label{fig:BGnoise}
\end{figure}

We measured the background noise three times, as mentioned in section \ref{sec:2}. In Figure \ref{fig:BGnoise} the background levels recorded at the fourth receiver are shown as recorded before the measurements (Pre), between the measurements with the J11 and the airgun (Mid) and after all measurements had been concluded (Post). The background noise level which we will show in the figures is the maximum of the measured signals (shown in grey in Figure \ref{fig:BGnoise}). The background measurements are similar for all receivers (not shown), however, due to the larger distance from the source the recorded signal levels are lower and the background level of the fourth receiver will be most limiting.

\section{Sound generated by the bubble curtain(s)} \label{app:BCnoise}
\begin{figure}[!ht]
\centering
\includegraphics[width=\columnwidth]{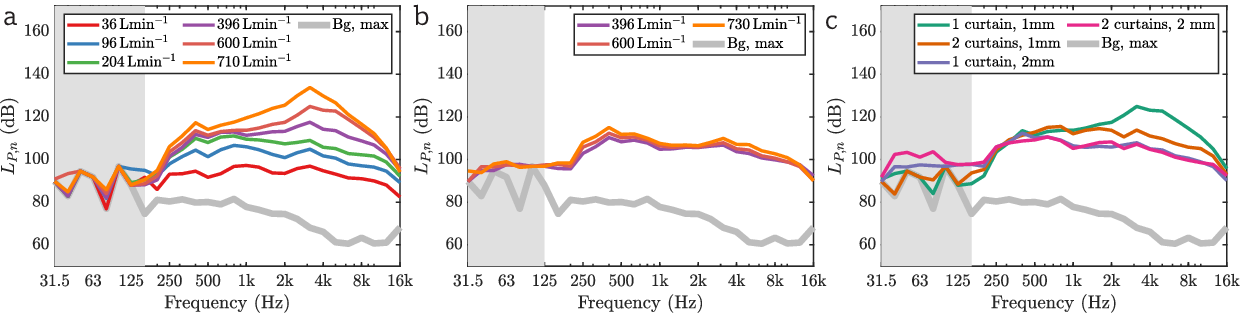}
\caption{(color online) Recorded sound from the bubble curtains for different flow rates. a) With 1 mm nozzles b) With 2 mm nozzles c) Comparison between the sound generated by different configurations at $600\,\mathrm{Lmin^{-1}}$} \label{fig:BubbleCurtainNoise}
\end{figure}

The bubble curtains generate noise themselves. We measured the generated sound for the different configurations we use. In Figure \ref{fig:BubbleCurtainNoise}a it is shown that the generated sound levels by one manifold with 1 mm nozzles is highly air flow rate dependent. Particularly at higher frequencies the sound pressure level increases with increasing air flow rate. For one manifold with 2 mm nozzles we only investigated the higher air flow rates, and found no clear flow rate dependency (see Figure \ref{fig:BubbleCurtainNoise}b). In Figure \ref{fig:BubbleCurtainNoise}c it is clear that two bubble curtains generated by 1 mm nozzles generate less sound than one with the same total air flow rate. The main reason being the halved air flow rate per manifold (see Figure \ref{fig:BubbleCurtainNoise}a) and the fact that the noise produced by the first manifold is shielded by the bubble curtain emitting from the second manifold. We see that the 2 mm nozzles generate generally less noise at higher frequencies ($>250\,\mathrm{Hz}$) compared to the 1 mm nozzles, but do the opposite at lower frequencies ($<250\,\mathrm{Hz}$). 

The sound generated by the bubble curtains is well above the background noise levels in the region of interest. This implies that the sound level by the bubble curtains can now be considered the background noise level for the measurements taken with a bubble curtain active. These lines are therefor generally shown in stead of the background noise levels. We furthermore present the 2 mm nozzle results if possible due to the slightly lower sound levels in the region of interest.

\section{Sound generated by the sources}\label{app:soundsource}
\begin{figure}[!ht]
\centering
\includegraphics[width=\columnwidth]{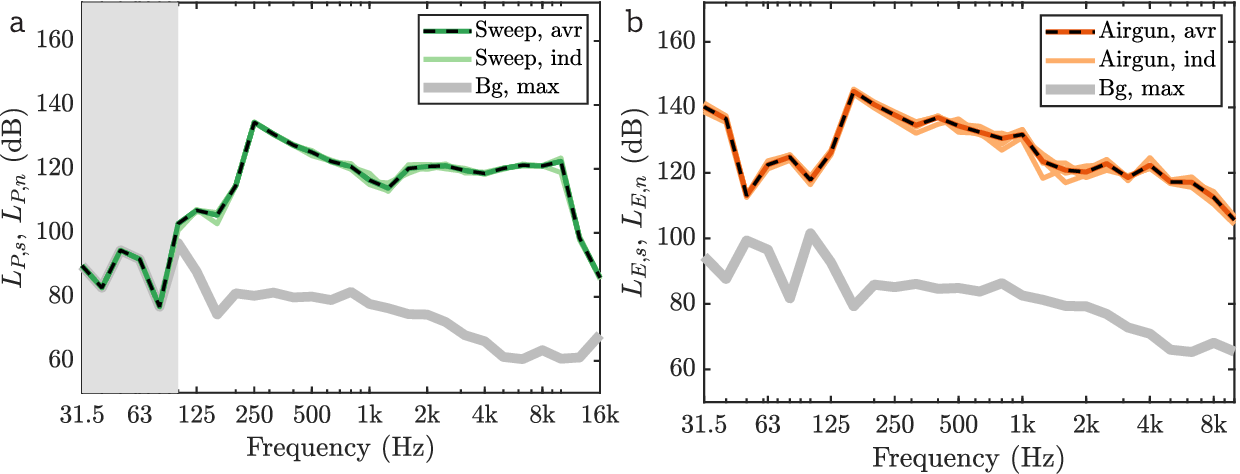}
\caption{(color online) a) Sound generated by the J11 hydro sounder measured by the fourth receiver. b) Sound generated by TNO's mini airgun measured by the fourth receiver} \label{fig:SourceSweepGun}
\end{figure}

The J11 hydro sounder is actuated by logarithmic sweeps in a range of $0.1-10\,\mathrm{kHz}$, however, the source generates additional, unintended, harmonics which can also be used in our analysis. For measuring the IL having harmonics is not a problem as long as the signal to noise ratio is sufficient it can be used. If the total noise level is less than 3 dB above the background noise level the source level can no longer be considered distinguishable, for that reason we disregard this part of the measurement (see the grey area and the dashed line in Figure \ref{fig:SourceSweepGun}a). We plot both the individual sweep results and the averaged result, showing that the signal is well reproducible. The mini airgun generates, mainly at the lower frequencies, a significantly higher sound level. At higher frequencies the signal generated by the airgun becomes less constant between different instances. Above $10\,\mathrm{kHz}$ we stop the analysis since the source level is limited and the spread in the signal yields unreliable results.

\section{Variations in the IL}\label{app:ILvar}

In this appendix we focus on estimating the contribution of the temporal variations in the plume on the variation in the IL. This requires the exclusion of the variation due to the airgun.

If we consider the following relation for the SEL at the fourth receiver based on the SEL at the first receiver without a bubble curtain:

\begin{equation}
    \mathrm{SEL}_{nc,4}=\mathrm{SEL}_{nc,1}-\Delta\mathrm{SEL}_{nc,1\rightarrow 4},
    \label{eq:Selnc4}
\end{equation}

with $\Delta\mathrm{SEL}_{nc,1\rightarrow 4}$ the difference in sound exposure level between both locations without an active bubble curtain. We can further write the SEL at the first receiver without a bubble curtain as

\begin{equation}
    \mathrm{SEL}_{nc,1}=\mathrm{SEL}_{c,1}-\Delta \mathrm{SEL}_{c-nc,1}.
    \label{eq:Selnc1}
\end{equation}

The IL at the fourth receiver, based on Eq. \ref{eq:Selnc4} and Eq. \ref{eq:Selnc1}, is given by

\begin{equation}
\begin{split}
    \mathrm{IL}_4=\mathrm{SEL}_{nc,4}-\mathrm{SEL}_{c,4}=\\ \mathrm{SEL}_{c,1}-\Delta \mathrm{SEL}_{c-nc,1}-\Delta\mathrm{SEL}_{nc,1\rightarrow 4} -\mathrm{SEL}_{c,4}.
    \end{split}
\end{equation}

Splitting $\mathrm{IL}_4$ up into its components allows us to discuss the origin of the variations in $\mathrm{IL}_4$.
Variations in $\Delta\mathrm{SEL}_{nc,1\rightarrow 4}$ do not depend on the variations in the bubble curtain. The variations in this parameter are assumed to be very limited since the path of transmission does not change. In other words, if we were to measure this variable we would measure the difference in SEL at receiver 1 and 4 without a bubble curtain present and we expect that to be very reproducible.\\
Variations in $\Delta \mathrm{SEL}_{c-nc,1}$ are likely significant and likely stem from temporal variations in the (first) bubble curtain. The reflecting sound on the (first) bubble curtain is known to impact the measured SEL at the first receiver, see Figure \ref{fig:Alllines}a, and the level of the reflected sound likely depends on the instantaneous bubble distribution. \\
So the variations in $\mathrm{SEL}_{c,1} -\mathrm{SEL}_{c,4}$ are likely due to the temporal variations of the bubble curtain and related to the variations in $\mathrm{IL}_4$. However we have now replaced the uncertainty of the variations in the source level with the uncertainty of the variations in $\Delta \mathrm{SEL}_{c-nc,1}$. The benefit of this approach is that the variations in the IL now only depend on the instantaneous state of the bubble curtain since these values are measured using the same instance of the source level. 

\clearpage
\section*{References}
\bibliography{Biblio}

\end{document}